%% file: main.tex
\documentclass[sn-mathphys]{sn-jnl}
\jyear{2022}
\pdfoutput=1
\expandafter\let\csname equation*\endcsname\relax
\expandafter\let\csname endequation*\endcsname\relax
\usepackage[english]{babel}
\usepackage[utf8]{inputenc}
\usepackage{amsmath,amssymb,amsfonts,latexsym}
\usepackage{graphicx}
\usepackage{mathrsfs} 
\usepackage{color}
\usepackage{booktabs}
\usepackage{bm}
\usepackage{braket}
\usepackage{nicefrac}
\usepackage{soul}
\usepackage{mathtools}
\usepackage{ulem}
\usepackage{dsfont}

\usepackage{amsmath,amssymb,amsfonts,latexsym}

\usepackage{braket}
\definecolor{blue(pigment)}{rgb}{0.2, 0.2, 0.6}
\usepackage{times}
\usepackage{hyperref}
\hypersetup{
    colorlinks=true,
    linkcolor=blue,
    citecolor=magenta,      
    urlcolor=magenta,
    pdftitle={epj-paper},
    pdfpagemode=FullScreen,
    }
\renewcommand*{\href}[2]{\\ #2,\\ {\bf url:}~\url{#1}}
\def\eqref#1{(\ref{#1})}

\newcommand{\bea}{\begin{eqnarray}}
\newcommand{\eea}{\end{eqnarray}}
\usepackage{setspace}
\renewcommand*{\geq}{\geqslant}
\renewcommand*{\leq}{\leqslant}
\newcommand{\I}{\ensuremath{\mathbf{i}}}
\include{new_commands}
\graphicspath{{Figures/}}
\usepackage{times}
\begin{document}

\title[Scaling of fronts and entanglement spreading during a domain wall melting]{\Large Scaling of fronts and entanglement spreading during a domain wall melting}

%\author*[2]{\fnm{Thierry} \sur{Platini}}
\author*[1]{\fnm{Stefano} \sur{Scopa}}\email{sscopa@sissa.it}
\author*[2]{\fnm{Dragi} \sur{Karevski}}\email{dragi.karevski@univ-lorraine.fr}
\affil[1]{SISSA and INFN, via Bonomea 265, 34136 Trieste, Italy}
%\affil*[2]{Statistical Physics Group, Centre for Fluid and Complex Systems, Coventry University, Coventry, England}
\affil[2]{Universit\'e de Lorraine, CNRS, LPCT, F-54000 Nancy, France}

\abstract{We revisit the out-of-equilibrium physics arising during the unitary evolution of a one-dimensional {XXZ} spin chain initially prepared in a domain wall state $\vert\psi_0\rangle=\vert\dots \uparrow\uparrow\downarrow\downarrow\dots\rangle$. In absence of interactions, we review the exact lattice calculation of several conserved quantities, including e.g. the magnetization and the spin current profiles. At large distances $x$ and times $t$, we show how these quantities allow for a ballistic scaling behavior in terms of the scaling variable $\zeta= x/t$, with exactly computable scaling functions. In such a limit of large space-time scales, we show that the asymptotic behavior of the system is suitably captured by the local occupation function of spinless fermionic modes, whose semi-classical evolution in phase space is given by a Euler hydrodynamic equation. Similarly, analytical results for the asymptotic fronts dynamics are obtained for the interacting chain via Generalized Hydrodynamics. In the last part of the work, we include large-scale quantum fluctuations on top of the semi-classical hydrodynamic background in the form of a conformal field theory that lives along the evolving Fermi contour. With this procedure, dubbed quantum generalized hydrodynamics, it is possible to obtain exact asymptotic results for the entanglement spreading during the melting dynamics.}

\maketitle
%{\small\tableofcontents}
%\hrulefill

\section{Introduction}\label{sec1}
In the last two decades or so, the study of transport phenomena in one-dimensional quantum systems is experiencing renewed interest due to the recently developed theoretical and experimental technologies. Special attention has been dedicated to one-dimensional integrable systems, where unusual transport properties emerge due to the presence of an infinite set of conserved charges. In particular, such conservation laws are responsible for a lack of thermalization of the system at large times and, consequently, for the emergence of current-carrying steady states, denoted as generalized Gibbs ensembles (GGE) \cite{Rigol2007}.\\

However, in the absence of translational invariance, the study of these transport properties becomes much more complicated as conventional integrability-based techniques are generically lost. In order to investigate transport phenomena in those systems with a non-flat density profile (resulting, for instance, from the presence of confining potentials), an Euler hydrodynamic theory for integrable models has been developed in recent years. This approach, dubbed Generalized Hydrodynamics (GHD) \cite{Bertini2016,Castro-Alvaredo2016} (see also Refs.~\cite{ghd-rev,ghd-notes,Essler-ghd-rev,DeNardis-rev} for reviews), allowed us to obtain asymptotically exact results in quantitative agreement with numerical simulations (e.g.~\cite{Bertini2016,Castro-Alvaredo2016,Collura2017,DeLuca2017,Bastianello2019,Bulchandani2017,Bulchandani2018,Doyon2017,Doyon2018,Piroli2017,Caux2019,Scopa2022}) and experimental data \cite{Schemmer2019,Malvania2020}.
GHD has been initially developed for models featuring an elementary Bethe ansatz, such as bosons with contact interactions \cite{ghd-JB} and spin chains \cite{Bertini2021}, and later extended over the last years to include multi-component models \cite{Nozawa2020,Nozawa2021,Mestyan2019,Scopa2021,Moller-2component,Scopa2022d}, diffusive corrections \cite{DeNardis2018,DeNardis2019,Medenjak2020,Durnin2021}, atom losses \cite{Bouchoule2020},  and even weak integrability-breaking terms \cite{Durnin2020,Bastianello2020b,Bastianello2021}. Recent developments have enabled the further study of intrinsically quantum properties such as zero-temperature entanglement and equal-time correlation functions, thanks to the requantization of GHD by means of a non-standard Luttinger liquid theory \cite{Ruggiero2019,Ruggiero2020,Collura2020,Scopa2021a,Scopa2022,Scopa2022b,Scopa2022c,Rottoli2022,Ruggiero2022}.\\

This success was made possible by pioneering studies conducted on so-called {\it bi-partite settings}, where two semi-infinite one-dimensional integrable models prepared in GGE states are attached together and let to unitarily evolve. At $t>0$, transport phenomena emerges from the junction of the two chains, which acts as a source of stable quasiparticles whose propagation give rise to non-vanishing currents throughout the system. Notice that these bi-partite quench protocols are placed midway between homogeneous quench problems and genuinely inhomogeneous systems, and hence constitute an ideal playground to develop and test new approaches to quantum transport. One of the simplest realizations of bi-partite quench protocol is the {\it domain wall} (DW) configuration of a spin chain, namely a state $\vert\psi_0\rangle=\vert\uparrow\dots\uparrow\downarrow\dots\downarrow\rangle$ realized by joining together two ferromagnets of opposite magnetization. Despite its simplicity, the time evolution of such a state shows non-trivial features of non-equilibrium dynamics and, for this reason, it has been the main character of a large number of studies, including stability analysis \cite{Gochev1977,Gochev1983,Yuan2007}, exact computations for the free case e.g. \cite{Antal1999,Tasaki1,Tasaki2,Araki2000,Ogata2002,Aschbacher2003,Antal2008,Doyon2015,Scopa2021,Dubail2017,Allegra2016,Karevski2002,Hunyadi2004,Rigol2004,Rigol2015,Platini2005,Platini2007,DeLuca2013,DeLuca2014,Viti2016,Eisler2018}, approximate and numerical results for  the interacting integrable e.g. \cite{Gobert2005,Calabrese2008,Zauner2012,Halimeh2014,Alba2014,Vicari2012,Sabetta2013,Biella2016,Bernard2016,Vidmar2017,Langmann2017,abf-19,bfpc-18,mbpc-18} and non-integrable e.g.\cite{Jesenko2011,Hauschild2016,Karrasch2013,Rakovszky2019,Lerose2020,Coppola2022} chain.\\

The scope of this short review is to provide a concise (and yet comprehensive) discussion of the physics of the DW melting by revisiting early results and modern GHD approaches to the problem, thus giving a handbook of the main available results and of their derivation. The paper is organized as follows. In Sec.~\ref{sec:model}, we introduce the model, the quench protocol, and the quantities that we wish to characterize during the dynamics. The contents are organized in two main sections: {\it i)}~Sec.~\ref{sec:transport} dedicated to the study of the profiles of conserved charges and currents by means of exact lattice calculations and hydrodynamics; {\it ii)}~Sec.~\ref{sec:QGHD}, where we employ a requantization of the hydrodynamic theory for the study of entanglement. In both sections, the study of the non-interacting case is treated separately for a better exposition. In Sec.~\ref{sec:conclusion}, we give a short summary and some concluding remarks.

\section{The domain wall melting problem}\label{sec:model}
We consider the one-dimensional XXZ model with Hamiltonian
\be\label{xxz-model}
\Ha=-\frac{J}{4}\sum_{j=-\nicefrac{N}{2}+1}^{\nicefrac{N}{2}-1} \left(\ssx_j\ssx_{j+1}+\ssy_j\ssy_{j+1}+\Delta \ssz_j\ssz_{j+1}\right)
\ee
where $N$ is the number of lattice sites (which we assume to be even for future convenience), $\hat\sigma_j^a$, $a={\rm x,y,z}$, denotes the standard Pauli operators acting on site $j$ and $\Delta$ is the anisotropy parameter. The hopping amplitude $J$ defines the overall energy scale and will be set to unity from hereafter. We shall focus on the gapless regime of the system $\vert\Delta\vert< 1$, {with the usual parametrization} $\Delta=\cos\gamma$. We do not consider the case $\vert\Delta\vert>1$ because it can be shown with energy arguments that initial domain wall configurations do not melt, see Refs.~\cite{Misguich2017,Misguich2019}. The case $\vert\Delta\vert=1$ is very peculiar  \cite{Ljubotina2017,Ilievski2018}  and therefore it will be also excluded in this short review.\\

\begin{figure}[t]
\centering
\includegraphics[width=.8\textwidth]{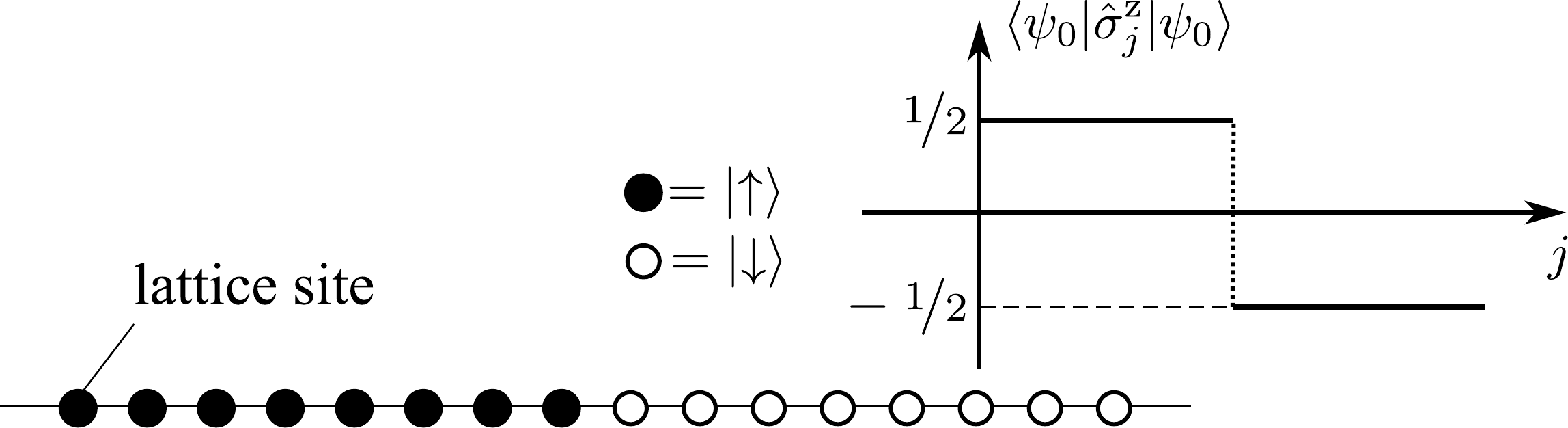}
\caption{Setup of the domain wall melting problem. At $t=0$ the spin chain is prepared in the product state \eqref{initial-state}, corresponding to a step-like profile of magnetization between the values $\pm1/2$. For $t>0$, this configuration is let to unitarily evolve with Hamiltonian \eqref{xxz-model}.}\label{fig:illustration_DW}
\end{figure}

At $t=0$, we prepare the system in the product state
\be\label{initial-state}
\vert\psi_0\rangle=\bigotimes_{j\leq0} \vert\uparrow_j\rangle \bigotimes_{j>0} \vert\downarrow_j\rangle
\ee
where  $\vert\uparrow_j\rangle$ (resp. $\vert\downarrow_j\rangle$) is the eigenstate of $\ssz_j$ with eigenvalue $+1$ (resp. $-1$), see Fig.~\ref{fig:illustration_DW} for an illustration. For $t> 0$ we consider the Hamiltonian dynamics generated by \eqref{xxz-model}
\be
\vert\psi_t\rangle=e^{- \I t \Ha}\vert\psi_0\rangle
\ee
during which the initial DW state \eqref{initial-state} gradually melts. During this process, we wish to characterize the transport properties of the system. To this end, we will investigate the non-equilibrium dynamics of the profiles of conserved charges and currents of the integrable model. For {the} sake of concreteness, we will highlight the main features related to the melting dynamics by considering the magnetization profile
\be
m_n(t)=\frac{1}{2}\langle\psi_t\vert \ssz_n\vert \psi_t\rangle,
\ee
and the spin current
\be
j^M_n(t)=\frac{1}{4}\langle\psi_t\vert \ssy_n\ssx_{n+1}-\ssx_n\ssy_{n+1}\vert\psi_t\rangle.
\ee
The latter {can be} deduced from the fact that the total magnetization is conserved by the unitary dynamics and, therefore,  a local magnetization current can be defined using a lattice continuity equation. {This extends to any conserved quantity and associated current in the model, which are all obtainable with simple hydrodynamic arguments as reviewed in the next section.}\\

 In the last part of this short review, we shall focus on the entanglement spreading. In particular, we will characterize the entanglement of a bipartition of the one-dimensional chain as $A\cup B\equiv [-\nicefrac{N}{2}+1,j]\cup[j+1,\nicefrac{N}{2}]$, for which the reduced density matrix is obtained as
\be
\hat\rho_A(t)={\rm tr}_B\vert\psi_t\rangle\langle\psi_t\vert
\ee
and, from it, the R\'enyi entropies are defined as
\be
S_{A}^{(\text{p})}(t)=\frac{1}{1-\text{p}}\log[{\rm tr}[(\hat\rho_A(t))^\text{p}]]
\ee
with index $\text{p}\in\mathbb{N}$. The entanglement entropy is then obtained in the limit $\text{p}\to 1$ and reads as
\be
S_A(t)\equiv \lim_{\text{p}\to 1}S^{(\text{p})}_A(t)=-{\rm tr}\hat\rho_A(t)\log\hat\rho_A(t).
\ee
%%%%%%%%%%%%%%%%%%%%%%%%%%
%%%%%%%%%%%%%%%%%%%%%%%%%%
\section{Transport properties during the DW melting}\label{sec:transport}
{As discussed below, the physics of the DW melting} is qualitatively similar in the whole region  $\vert\Delta\vert <1$. {However, we find convenient to discuss} the non-interacting case $\Delta=0$  separately since, in this case,
 the system reduces to a set of free fermions and the transport properties can be determined with exact lattice calculations.  
From this, an emergent hydrodynamic picture can be rigorously determined. In the interacting case instead, the characterization of the transport during the DW melting is possible only by means of GHD and it requires the review of the Bethe ansatz solution of the interacting model.
\subsection{Non-interacting case}\label{sec:transport-nonint}
For $\Delta=0$, the Hamiltonian \eqref{xxz-model} reduces to the XX model. The latter can be mapped to a system of spinless non-interacting fermions with tight-binding Hamiltonian
\be\label{xx-model}\begin{split}
\Ha_\text{xx}=-\frac{1}{4}\sum_{j=-\nicefrac{N}{2}+1}^{\nicefrac{N}{2}-1} (\ssx_j\ssx_{j+1}+\ssy_j\ssy_{j+1})=-\frac{1}{4}\sum_{j=-\nicefrac{N}{2}+1}^{\nicefrac{N}{2}-1} (\hat{c}^\dagger_j\hat{c}_{j+1}+\hat{c}_j\hat{c}^\dagger_{j+1}),
\end{split}
\ee
up to a boundary term which is negligible when $N\to\infty$, as considered below. Here $\hat{c}^\dagger_j$, $\hat{c}_j$ stand for the creation and annihilation operators of a lattice spinless fermion at site $j$ and they satisfy canonical anticommutation relations $\{\hat{c}^\dagger_i,\hat{c}_j\}=\delta_{i,j}$. The two forms of the Hamiltonian \eqref{xx-model} are related through Jordan-Wigner transformation \cite{Jordan1928,Wigner1997}
\be
\hat{c}^\dagger_j=\exp[\I \pi\sum_{i<j} \ssp_i\ssm_i] \ \ssp_j; \qquad \hat{c}_j=\exp[-\I \pi\sum_{i<j} \ssp_i\ssm_i]\ \ssm_j,
\ee
where $\hat\sigma^\pm_j$ are the ladder combination of Pauli operators $\hat\sigma^\pm_j=(\ssx_j\pm\I\ssy_j)/2$. In Fourier space, the model \eqref{xx-model} is readily diagonalized
\be\label{eq:free-fermions}
\Ha_{\rm xx}=-\sum_{q=1}^N \cos(k_q) \hat{b}^\dagger_q \hat{b}_q +\text{cst}
\ee
 in terms of the operators
\be
\hat{b}^\dagger_q=\sum_{j=-\nicefrac{N}{2}+1}^{\nicefrac{N}{2}} \chi_{q,j} \ \hat{c}^\dagger_j, \qquad \hat{b}_q=(\hat{b}^\dagger_q)^\dagger, \qquad \{\hat{b}^\dagger_q,\hat{b}_{q'}\}=\delta_{q,q'}
\ee
with single-particle wavefunction $\chi_{q,j}=\sqrt{2N^{-1}}\sin(k_q j)$ and quantized momenta $k_q=\pi q/N$, $q=1,\dots,N$.\\

%We further consider from hereafter
Taking the limit $N\to \infty$,  it is possible to replace the quantized momentum $k_q$ with a continuous variable $k\in[-\pi,\pi]$ and use the single-particle wavefunctions  $\chi_{k,j}=\exp(\I k j)/\sqrt{N}$.
%, since the Dirichlet boundary conditions at sites $j+\nicefrac{N}{2}=1,N$ are now lifted.
%%%%%%%%%%%%%%%%%%%%%%%%%%%%%%%%
\subsubsection{Lattice description of the dynamics}
%%%%%%%%%%%%%%%%%%%%%%%%%%%%
\begin{figure}[t]
\centering
\includegraphics[width=.75\textwidth]{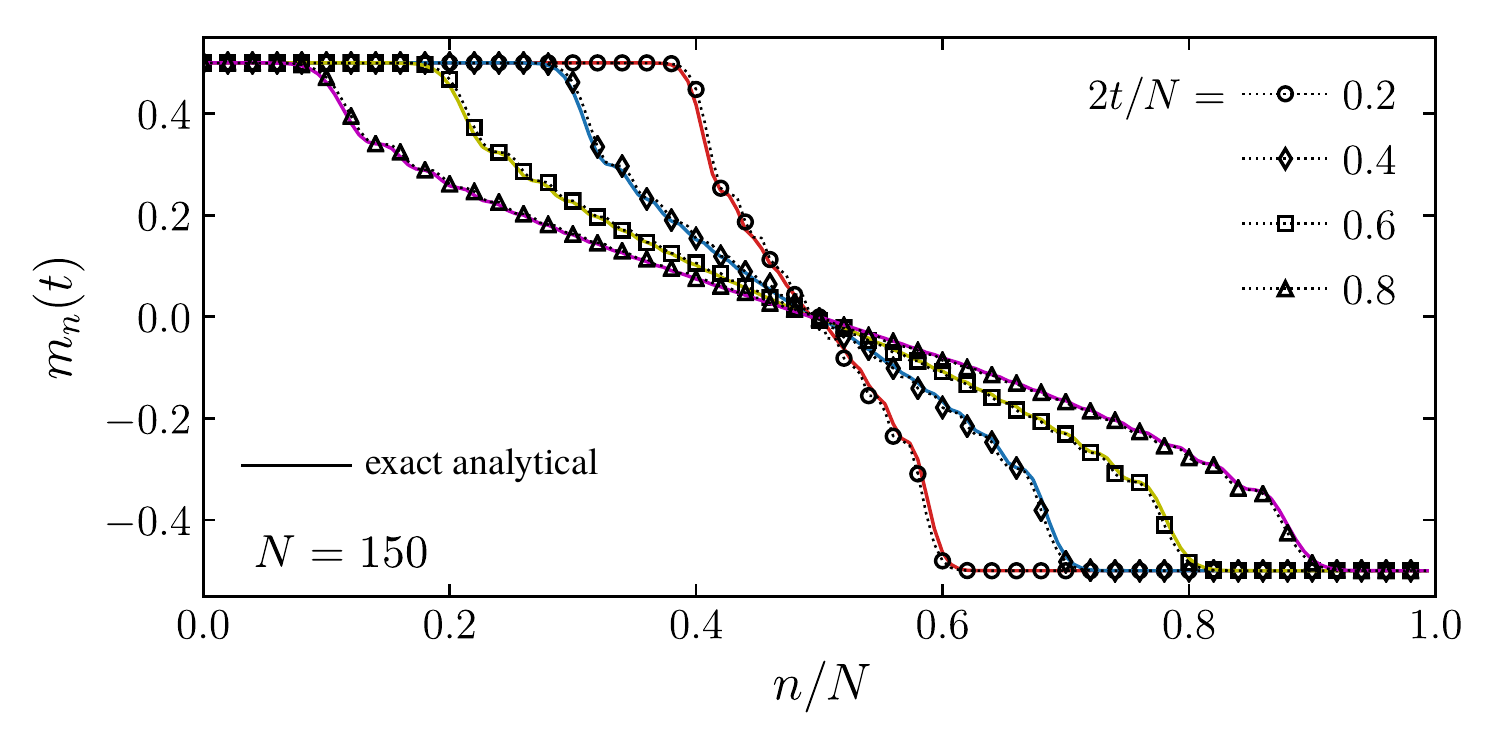}
\caption{Magnetization profiles as function of the lattice site $j$, plotted at different times. The symbols show the numerical data, obtained from the exact diagonalization of the lattice Hamiltonian with $N$ sites and by a further Trotter evolution of the initial two-point function, see e.g. Refs.~\cite{Scopa2021a,Scopa2022b} for details. The full lines show the exact result of Eq.~\eqref{magn-lattice}.% Notice that the major contribution to Eq.~\eqref{magn-lattice} comes from the ${\cal J}_\nu$ with small $\nu$, see e.g. Ref.~\cite{Antal1999}. Indeed, we observe a very fast convergence of the analytical curve already after considering a few terms in the sum of Eq.~\eqref{magn-lattice}.
}\label{fig:lattice-magn}
\end{figure}
%%%%%%%%%%%%%%%%%%%%%%%%%%%%
From the exact solution of the non-interacting spin chain \eqref{eq:free-fermions}, one can determine the time evolution of the operators $\hat{b}^\dagger_k$ in Heisenberg picture as
\be
\hat{b}^\dagger_k(t)=\exp[-\I t \cos(k)] \ \hat{b}^\dagger_k(0)
\ee
and obtain, via inverse Fourier transform, the time evolution of the lattice fermionic operators as
\be
\hat{c}^\dagger_n(t)=\sum_{j\in\mathbb{Z}} \I^{n-j} {\cal J}_{j-n}(t) \ \hat{c}^\dagger_j(0),
\ee
where ${\cal J}_\nu(z)$ are Bessel functions of the first kind. From this observation, the expectation value of lattice operators can be obtained by elementary calculations. For instance, the magnetization profile is given by
\be\label{magn-lattice}
m_n(t)=\langle \psi_0\vert \ssz_n(t)\vert \psi_0\rangle=\sum_{j,l\in\mathbb{Z}} \I^{j-l} {\cal J}_{j-n}(t) {\cal J}_{l-n}(t) \langle\psi_0\vert\hat{c}^\dagger_l(0)\hat{c}_j(0)\vert\psi_0\rangle -\frac{1}{2}.
\ee
The initial domain wall state \eqref{initial-state} forces $j\equiv l$  in the previous double sum leading to \cite{Schutz1999,Antal1999,Karevski2002,Platini2005,Antal2008}
\be
m_n(t)=-\frac{1}{2}+{\sum_{j\leq -n} }{\cal J}_j^2(t) = -\frac{1}{2} \sum_{j=-n+1}^{n-1} {\cal J}_j^2(t),
\ee
where the last expression is obtained using the properties ${\cal J}_{-k}(z)= (-1)^k{\cal J}_k(z)$ and $\sum_{k\in\mathbb{Z}}{\cal J}_{k}^2(z) =1  $. In Fig.~\ref{fig:lattice-magn}, we compare the exact result in Eq.~\eqref{magn-lattice} with numerical calculations for the {finite} lattice model, finding an excellent agreement.\\

In general, given the non-interacting nature of the problem, one can determine exact results for the transport properties during the DW melting from the knowledge of the two-point function
\be
G_{n,m}(t)\equiv\langle \psi_0\vert \hat{c}^\dagger_n(t)\hat{c}_m(t)\vert \psi_0\rangle=(-\I)^{n-m}\sum_{s=0}^\infty {\cal J}_{s+n}(t){\cal J}_{s+m}(t),
\ee
{which can be simplified, thanks to the summation theorem \cite{Gradshteynp989}  for Bessel functions, to \cite{Schutz1999,Antal1999,Karevski2002,Platini2005,Antal2008}
\be\label{eq:two-point-func}
G_{n,m}(t)=(-\I)^{n-m} [{\cal J}_n(t){\cal J}_m(t) +{\frac{t}{2 (n-m)}}({\cal J}_n(t){\cal J}_{m+1}(t)-{\cal J}_{n+1}(t){\cal J}_m(t))].
\ee
For instance, the magnetization profile in Eq.~\eqref{magn-lattice} is recovered from the diagonal of $G_{n,m}$ as
\be
m_n(t)=-\frac{1}{2}+  G_{n,n}(t)
\ee
and the spin current can be obtained as
\be\label{jM-lattice}
j_n^M(t)=\frac{1}{2\I}[G_{n,n+1}(t)-G_{n+1,n}(t)].
\ee
Similar results for other quantities can be worked out following the same lines.

\subsubsection{Scaling limit of fronts}
In the limit $n\to \infty$, $t\to\infty$ at fixed ratio $\zeta=n/t$, the magnetization profile admits a scaling form 
\be
m_n(t)\sim \Phi(n/t)
\ee
with a scaling function $\Phi(\zeta)$ that can be analytically determined from Eq.~\eqref{magn-lattice} using the asymptotic expansion of ${\cal J}_n(n/\zeta)$ when $n\to \infty$. {From this calculation, one obtains
\be\label{scaling-magn}
\Phi(\zeta)=\begin{cases} 
-1, \quad \text{if $\zeta<-1$};\\[3pt]
-\frac{1}{\pi}\arcsin(\zeta), \quad \text{if $-1\leq \zeta\leq 1$};\\[3pt]
0, \quad \text{if $\zeta>1$}
\end{cases}
\ee
Eq.~\eqref{scaling-magn} highlights the presence of a non-equilibrium steady state (NESS) for this quench protocol, characterized by a non-homogeneous profile of magnetization. Since $\ssz_j$ is a conserved quantity for the spin chain \eqref{xxz-model}, it follows that the NESS is characterized by the presence of a non-vanishing spin current $j^M_n(t)$ that can be determined from {the continuity equation}
\be\label{cons-law}
\de_t m_n(t) +\nabla_n \ j^M_n(t) =0
\ee
where $\nabla_n$ is a discrete derivative defined on the lattice. Using Eq.~\eqref{scaling-magn} in \eqref{cons-law}, one finds the scaling form for the NESS spin current
\be\label{asy-current}
j^M_n(t)\sim \Xi(\zeta)=\begin{cases}
\frac{1}{\pi}\sqrt{1-\zeta^2}, \quad \text{if $-1\leq\zeta\leq 1$};\\[3pt]
0, \quad \text{otherwise}.
\end{cases}
\ee
The same result is obtained from Eq.~\eqref{jM-lattice} by taking the asymptotic behavior of the Bessel functions.

\subsubsection{Emergent hydrodynamic description}\label{hydro}
In the limit $N\to\infty$ and for a translationally-invariant Fermi gas, the occupation function of fermionic modes can be obtained from the Fourier transform of the two-point correlation function in real space
\be
n_k\equiv\langle\hat{b}^\dagger_k\hat{b}_k\rangle =\sum_{n\in\mathbb{Z}} e^{-\I k n} \langle\hat{c}^\dagger_{j+n}\hat{c}_j\rangle,
\ee
where the site $j$ is arbitrary because of translational invariance. Given the non-interacting nature of the problem, $n_k$ is expected to fully characterize the properties of the fermionic gas. In order to investigate the DW melting problem then, one can consider $j=\zeta t$ and take the limit $t\to \infty$ before performing the summation over the index $n$ \cite{Antal1999}. In this way, the local properties of the fermionic gas are characterized by the steady-state value of a region which is shifted from the orgin of an amount $\zeta t$. Using the expression \eqref{eq:two-point-func} for the two-point function, and taking the asymptotic expansion of the Bessel functions, one arrives at the expression
\be
\lim_{t\to\infty} \langle\psi_0\vert\hat{c}^\dagger_{\zeta t+n}(t)\hat{c}_{\zeta t}(t)\vert \psi_0\rangle=\frac{\I^n}{\pi n}\sin[\arccos(\zeta)n]=\frac{1}{2\pi}\int_{\arcsin(\zeta)}^{\pi-\arcsin(\zeta)} \dd p \ e^{\I p n}
\ee
and after a Fourier transform, it finally leads to \cite{Antal1999}
\be\label{n-asy-bessel}
n_k(\zeta\equiv\nicefrac{j}{t})=\frac{1}{2\pi}\int_{\arcsin(\zeta)}^{\pi-\arcsin(\zeta)} \dd p \sum_{n\in\mathbb{Z}} e^{\I (p-k) n}=
{\frac{1}{2\pi}\int^{\pi-\arcsin(\zeta)}_{\arcsin(\zeta)} \dd p \ \delta(k-p).}
\ee
We argue that this exact asymptotic result for the occupation function $n_k(\zeta)$ allows for an intuitive hydrodynamic interpretation. Indeed  consider again the tight-binding fermionic Hamiltonian \eqref{xx-model}. The continuum limit is then attained by dividing the chain into a set of equally-spaced interval of size $\Delta  x=Ma$, where $a$ is the lattice spacing and $M\gg 1$ is the number of sites in each interval $\Delta x$. By taking the scaling limit where $a,\Delta x\to 0$ at fixed $\Delta x/a=M$, the lattice site $j\in\mathbb{Z}$ is replaced by the continuous variable $x\equiv a j\in\mathbb{R}$, and the Hamiltonian \eqref{xx-model} is written as \cite{Wendenbaum2013,Scopa2021a}
\be\label{eq:xx-continuous}
\Ha_\text{xx}=\int_{-\infty}^\infty \dd x \ \int_0^{\Delta x} \frac{\dd y}{a\Delta x} [-\frac{1}{2}(\hat{c}^\dagger_{x+y}\hat{c}_{x+y-a} +\text{h.c.})]
\ee
with continuous fermionic operators $\hat{c}^\dagger_x\equiv \hat{c}^\dagger_{ja}=\hat{c}_j$, $\hat{c}_x=(\hat{c}^\dagger_x)^\dagger$. Furthermore, around each coarse-grained point, the system is assumed to be in an eigenstate of a system in a periodic box of size $\Delta x$. Therefore, the Hamiltonian \eqref{eq:xx-continuous} can be diagonalized in Fourier space as
\be
\Ha_\text{xx}=\int_{-\infty}^\infty \dd x  \int_{-\pi}^\pi \frac{\dd k}{2\pi} [-\cos(k) ] \hat{b}^\dagger_{k,x}\hat{b}_{k,x},
\ee
where now the local excitations in $x$ are created by $\hat{b}^\dagger_{k,x}$, defined through
\be
\hat{c}^\dagger_{x+y}=\int_{-\pi}^\pi \frac{\dd k}{2\pi} \ e^{\I k y} \ \hat{b}^\dagger_{k,x}, \quad (\hat{b}^\dagger_{k,x})^\dagger\equiv \hat{b}_{k,x}.
\ee 
At this point, one can define the local fermionic occupation function for the DW initial state as 
\be\label{Wigner-func}
n_k(x)=\int_0^{\Delta x} \frac{\dd y}{a\Delta x} \int_{-\infty}^\infty \dd z \ e^{\I k z} \ \langle \psi_0\vert \hat{c}^\dagger_{\nicefrac{x+y+z}{2}}\hat{c}_{\nicefrac{x+y-z}{2}}\vert\psi_0\rangle.
\ee
For free fermionic systems, one can notice from Eq.~\eqref{Wigner-func} that the local occupation function of fermionic modes corresponds to the Wigner function of the system. In the hydrodynamic limit that we are considering, the Wigner function takes only positive values $n_k(x)\in[0,1]$ and it carries the physical interpretation of probability of finding a particle of momentum in a coarse-grained position $(x,k)$ of the phase space. It is then easy to see that the macrostate corresponding to the initial state \eqref{initial-state} is given by
\be\label{eq:n0}
n_k(x)=\begin{cases}
1, \quad\text{if $x\leq0$ and $-\pi \leq k \leq \pi$};\\[3pt]
0, \quad\text{otherwise.}
\end{cases}
\ee
%%%%%%%%%%%%%%%%%%%%%%%%%%%%%%%%%
\begin{figure}[t]
\centering
\includegraphics[width=\textwidth]{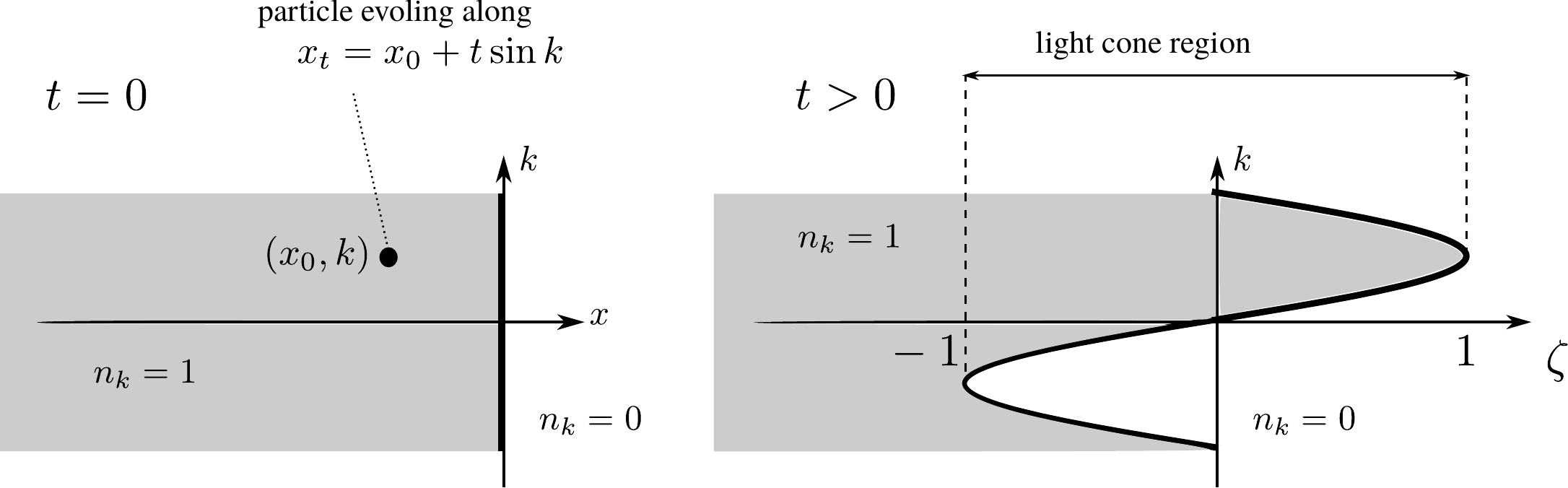}
\caption{Hydrodynamic description of the DW melting problem for the non-interacting spin chain. Left~--~ The initial state \eqref{initial-state} can be described in terms of the local fermionic occupation function \eqref{eq:n0}, here illustrated in the position-momentum plane, i.e., in the phase-space of the free Fermi gas. Right~--~ At times $t>0$, the evolution of the occupation function is obtained from the equation of motion of each non-interacting mode and it displays a non-trivial phase-space configuration in the region $\vert x\vert\leq t$ (light cone region), where the correlations during the DW melting process are spread from the junction, see main text.}\label{fig:sketch-n-free}
\end{figure}
%%%%%%%%%%%%%%%%%%%%%%%%%%%%%%%%%
The form of Eq.~\eqref{eq:n0} for the hydrodynamic description of the DW state is self-evident as it fills the l.h.s. of the system with modes $-\pi\leq k\leq \pi$, leaving the right part empty. Due to the non-interacting nature of the particles, each mode of momentum $k$, initially at position $x_0$, evolves ballistically with constant velocity $v(k)=\sin k$ according to the equation of motion
\be\label{eq:trajectory}
x_t= x_0 + v(k)t .
\ee
It follows that the time-evolved fermionic occupation function takes the form
\be
n_k(x,t)=n_k(x-v(k) t, 0).
\ee
The latter can be also interpreted as the solution of the following Euler equation
\be\label{Moyal}
(\de_t + v(k) \de_x)n_k(x,t)=0,
\ee
which is satisfied by the Wigner function at lowest order in $\de_x$ and $\de_k$ derivatives. A careful derivation of Eq.~\eqref{Moyal} can be found in Refs.~\cite{Fagotti2017,Fagotti2020}. We illustrate this hydrodynamic evolution in Fig.~\ref{fig:sketch-n-free}. \\

Moreover, since our initial state \eqref{eq:n0} has zero entropy and since the Euler equation \eqref{Moyal} is known to preserve the entropy of a given initial state at any time during the time evolution, the solution for $n_k(x,t)$ can be reconstructed by following the time evolution of its contour. For intuitive reasons, the latter is typically referred to as Fermi contour $\Gamma_t$ and it contains the information about the local Fermi points $k_F^\pm$ of the model at any space-time position $(x,t)$ \cite{Scopa2022,Scopa2022b}:
\be\label{Fermi-contour}
\Gamma_t=\bigcup_x [k_F^-(x,t), k_F^+(x,t)].
\ee
By noticing that the Fermi contour in the initial state is given by those particle living at the interface at $x=0$, the expression for $k^\pm_F(x,t)$ is easily obtained from Eq.~\eqref{eq:trajectory} as
\be
k^\pm_F(x,t)=\left\{\pi-\arcsin(\nicefrac{x}{t}); \ \arcsin(\nicefrac{x}{t})\right\}, \quad \text{if $\vert\nicefrac{x}{t}\vert<1$}.
\ee
Using this, one can write the time-evolved fermionic occupation function as
\be\label{n-t-nonint}
n_k(x,t) \equiv n_k(\zeta=\nicefrac{x}{t})=\begin{cases} 1,\quad\text{if $k_F^-(\zeta)\leq k\leq k_F^+(\zeta)$};\\[3pt]
0,\quad \text{otherwise}\end{cases} ,
\ee
hence recovering the result in Eq.~\eqref{n-asy-bessel} via hydrodynamic arguments. From Eq.~\eqref{n-t-nonint} one can obtain the asymptotic behavior of conserved charges and associated currents straightforwardly. Indeed, given a local conserved charge $\hat{\cal Q}$ of the model \eqref{xx-model}, the expectation value of the charge density is given as the integral sum over the available modes at space-time position $(x,t)$
\be\label{all-charges}
q_x(t)\equiv \nicefrac{\langle \hat{\cal Q}\rangle}{N}=\int_{-\pi}^\pi\frac{\dd k}{2\pi} n_k(\nicefrac{x}{t}) \ h^{({\cal Q})}_k=\int_{k_F^-(\zeta)}^{k_F^+(\zeta)}\frac{\dd k}{2\pi} \ h^{({\cal Q})}_k,
\ee 
weighted with the single-particle eigenvalue $h^{({\cal Q})}_k$ associated with $\hat{\cal Q}$. Similarly, the associated current is obtained as
\be\label{all-currents}
j^{{\cal Q}}_x(t)=\int_{k_F^-(\zeta)}^{k_F^+(\zeta)}\frac{\dd k}{2\pi} \ v(k) \ h^{({\cal Q})}_k.
\ee 
%%%%%%%%%%%%%%%%%%%%%%%%%%%%%%%%%
\begin{figure}[t]
\centering
\includegraphics[width=\textwidth]{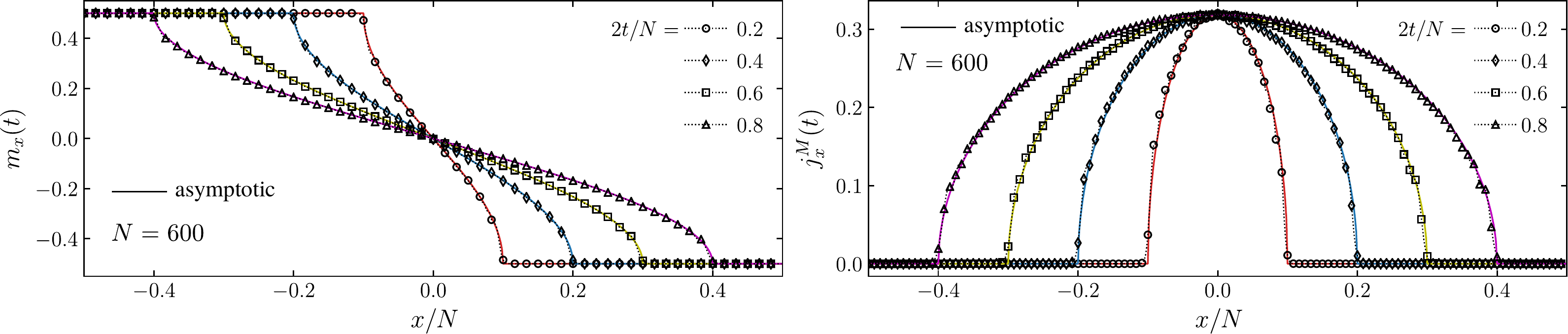}
\includegraphics[width=\textwidth]{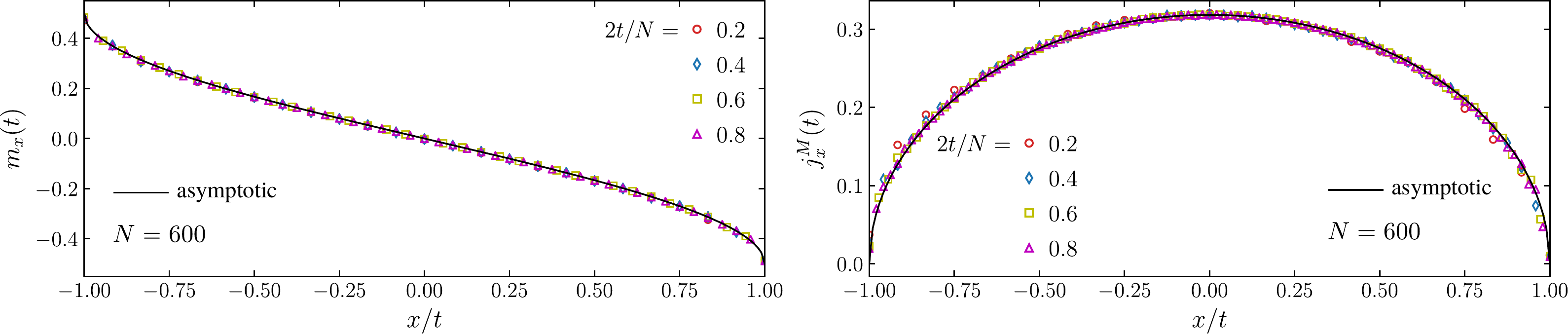}
\caption{Asymptotic behavior of the charges and current profiles during the DW melting for the non-interacting spin chain. Top panels~--~Asymptotic magnetization (left) and spin current (right) profiles as function of the rescaled position $x/N$, for different values of time. The symbols show the numerical data obtained for a chain of $N\gg 1$ via exact diagonalization and subsequent Trotter evolution of the two-point function (see also the caption of Fig.~\ref{fig:lattice-magn}). The full lines show the scaling function in Eq.~\eqref{scaling-magn} (resp.~Eq.\eqref{asy-current}) for the magnetization (resp.~the spin current). Bottom panels~--~ Same plots as function of $\zeta=\nicefrac{x}{t}$. The data collapse at different times signals the presence of ballistic transport in our quench setting. }\label{fig:Hydro-M-jM}
\end{figure}
%%%%%%%%%%%%%%%%%%%%%%%%%%%%%%%%%
For instance, by setting $h^{({\cal Q})}_k\equiv h^{(M)}_k=1$ one can recover the fermionic density $n_x(t)\equiv m_x(t)+1/2$ and the associated current in Eqs.~\eqref{scaling-magn}-\eqref{asy-current}. Notice that using Eq.~\eqref{all-charges}-\eqref{all-currents}, the conservation law for each conserved quantity
\be
\de_t \ q_x(t) + \de_x \ j^{{\cal Q}}_x(t)=0
\ee
follows directly from the evolution of the filling function in Eq.~\eqref{Moyal}. In Fig.~\ref{fig:Hydro-M-jM}, we compare the asymptotic results for the profiles of $m_x(t)$ and $j^M_x(t)$ with exact numerical calculations for a spin chain of $N=600$ sites, finding an excellent agreement and a perfect data collapse in the scaling variable $\zeta=\nicefrac{x}{t}$.\\

Before turning to the analysis of the interacting spin chain, we wish to conclude this paragraph with a final remark. From both analytical and numerical calculations, we clearly observe a dependence of the asymptotic fronts on the scaling variable $\zeta=\nicefrac{x}{t}$, rather than on $x$ and $t$ separately. This feature is related to the Euler hydrodynamic description of the DW melting problem, and for integrable models, it gives rise to ballistic transport. Along a ray of fixed $\zeta$, the system keeps its steady-state configuration during the whole time evolution. For $\zeta\to\pm\infty$ (correspoding to either $\vert x\vert\to \infty$ or $t\to 0$), one inevitably moves outside the correlated region and hence recovers the initial DW configuration of the spin chain. Indeed, in our problem, the modes that are responsible for the non-equilibrium dynamics are originated at the junction $x=0$ and spread throughout the left and right part of the spin chain with a finite velocity, bounded by the maximum value $\max_k[v(k)]=1$.
Therefore, the system develops a non-trivial profile in the so-called {\it light cone region} $-t\leq x\leq t$, which is determined by the equation of motion of the fastest excitations $k=\pm\pi/2$, while it keeps its original configuration in those space-time positions where the propagating modes have not arrived yet. In general, a light-cone effect is expected whenever a maximal velocity of propagation exists (e.g. due to the Lieb-Robinson bound \cite{LR-bound}). We show this by plotting the profile of magnetization \eqref{scaling-magn} in a $(x,t)$ plane, see Fig.~\ref{fig:NESS}.
%%%%%%%%%%%%%%%%%%%%%%%%%%%%%%%
\begin{figure}[t]
\centering
\includegraphics[width=.8\textwidth]{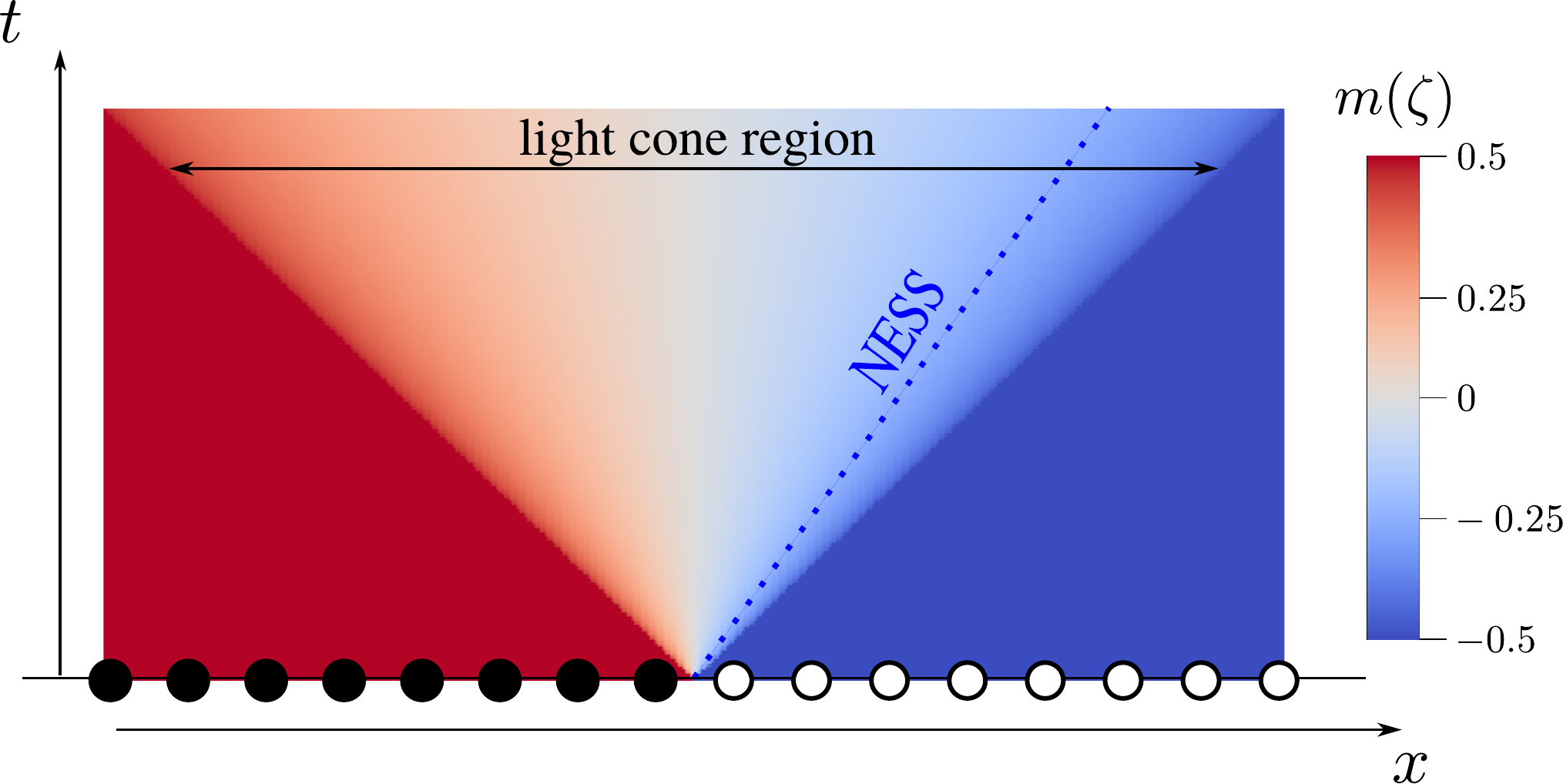}
\caption{The DW melting problem in the $x$-$t$ plane. The initial junction at $x=0$ acts as a source of propating particles, leading to the gradual melting of the initial ordered chain inside the light cone region $-t\leq x \leq t$. Outside the correlated region, the system keeps its initial DW configuration. Each ray of constant $\zeta=\nicefrac{x}{t}$ corresponds to a specific NESS for the dynamics (as shown in the figure by the uniform coloring of rays inside the light cone). We illustrate this feature by plotting the magnetization profile \eqref{scaling-magn} as a color plot .}\label{fig:NESS}
\end{figure}
%%%%%%%%%%%%%%%%%%%%%%%%%%%%%%%%%
\subsection{Interacting case}
We now turn to the analysis of transport during the DW melting of the interacting spin chain \eqref{xxz-model} , which we re-write for convenience
\be\label{xxz-model2}
\Ha=-\frac{1}{4}\sum_{j=-\nicefrac{N}{2}+1}^{\nicefrac{N}{2}-1} \left(\ssx_j\ssx_{j+1}+\ssy_j\ssy_{j+1}+\Delta \ssz_j\ssz_{j+1}\right).
\ee
As anticipated, we are interested in the regime $\vert\Delta\vert<1$, where it is customary to write $\Delta=\cos(\gamma)$. We further focus on the rational case, i.e., on those values of $\gamma$ that can be written as ratio $\gamma=\pi Q/P$ with $Q$, $P$ two-coprime integers $1\leq Q<P$. In this case, the interaction $\gamma$ admits a continued fraction representation
\be
\gamma = \frac{\pi}{\nu_1 + \frac{1}{\nu_2 + \frac{1}{\nu_3 + \dots}}},
\ee
where $\{\nu_1,\dots, \nu_q\}$ is a set of numbers satisfying $\nu_1,\dots,\nu_{q-1}\geq 1$ and $\nu_q\geq 2$. In the limit $N\to\infty$, the model \eqref{xxz-model2} is exactly solved by means of Thermodynamic Bethe Ansatz (TBA), see e.g. Refs.~\cite{Takahashi-book,Korepin2010}. In particular, for large $N$ and according to the string hypothesis \cite{Takahashi-book}, the excitation spectrum of the spin chain is described by different species of quasiparticles, generically referred as {\it strings}. {The total number $\delta$ of  strings species  is determined by the interaction parameter $\gamma$ through }\cite{Collura2017,Scopa2022}
\be
\delta=\sum_{i=1}^q \nu_i.
\ee
{As we will see in the following paragraphs, the asymptotic physics of the DW melting in the interacting case is found to be qualitatively similar to the non-interacting one. }
%In the presence of interactions one needs to replace the various hydrodynamic quantities with the corresponding ones for the interacting quasiparticles.}

\subsubsection{Thermodynamic Bethe ansatz solution}
In the following, we provide a short summary of the TBA solution of the model \eqref{xxz-model2} tailored for the introduction of the objects which are needed for the DW melting problem discussed below. The interested reader can find a comprehensive treatment e.g.~in Refs.~\cite{Takahashi-book,Korepin2010}.\\

In the presence of interaction $\Delta\neq0$, the diagonalization of the spin chain \eqref{xxz-model2} can be performed via Bethe ansatz. In particular, in the Hilbert space sector with $M$ spins up and assuming periodic boundary condition for the spin chain, one can write down the exact eigenstates of the model in terms of some complex parameters $\{\lambda_j\}_{j=1}^M$ (usually called {\it rapidities}), which generalize the concept of particle momenta of a free Fermi gas to the interacting case. {The rapidities $\lambda_j$ are obtained through the solution of non-linear algebraic equations implementing non-trivial quantization condition for the interacting model, see e.g. Refs.~\cite{Takahashi-book,Collura2017,Scopa2022}.} \\
As  mentioned above, for sufficiently large $N$, the rapidities of the model are arranged in symmetric patterns around the real axis  called strings ${\cal S}_j$
\be
{\cal S}_j=\lambda_j^\alpha + \frac{\I\gamma}{2} F_j, \quad j=1,\dots,\delta.
\ee
Each string specifies a quasiparticle species in the model. The parameter $\lambda_j^\alpha\in\mathbb{R}$ is called string center and the index $\alpha=1,\dots , \ell_j$ identifies a specific quasiparticle belonging to the string ${\cal S}_j$. The number $\ell_j$ determines the total number of quasiparticles of a given species. The quantities $\ell_j$ and $F_j$ can be determined using Bethe ansatz. Their precise expressions can be read in e.g.~Refs.~\cite{Takahashi-book,Collura2017,Scopa2022}. \\

In the limit $N\to\infty$ the spectrum of the model becomes densely populated and the state of the system can be suitably described in terms of a spectral distribution of quasiparticles 
\be
\rho^{(j)}(\lambda)\equiv \lim_{N\to\infty} \rho^{(j)}(\lambda_j^\alpha)= \lim_{N\to\infty} [N\vert \lambda^{\alpha+1}_j-\lambda^{\alpha}_j\vert]^{-1},
\ee
determined by the string center only. The spectral function $\rho^{(j)}(\lambda)$ can be obtained directly from the solution of the integral equation
\be
s_j \rho^{(j)}(\lambda) =a_j(\lambda) - \sum_{i=1}^\delta \int_{\infty}^\infty \dd \lambda' \ T_{j,i}(\lambda-\lambda') \ \rho^{(i)}(\lambda'),
\ee
where $s_j=\pm1$ is called string sign and $a_j(x)$, $T_{j,i}(x)$ are interaction kernels, see e.g.~Refs.~\cite{Takahashi-book,Collura2017,Scopa2022}~ for their expressions. The knowledge of $\rho^{(j)}(\lambda)$ fully characterize the thermodynamic properties of the zero-entropic states of the spin chain \eqref{xxz-model2}. For later purposes, we introduce also the occupation function for zero-entropic states as
\be
n^{(j)}(\lambda)=\begin{cases}1, \qquad \text{if $\rho^{(j)}(\lambda)\neq 0$};\\[3pt]
0, \qquad\text{otherwise.}
\end{cases}
\ee
For more generic states having a non-zero entropy (e.g.~a thermal state), one has to introduce a density distribution for the unoccupied rapidities (or {\it holes}) $\rho_h^{(j)}(\lambda)$ and complement the Bethe ansatz solution with additional thermodynamic arguments that relate the functions $\rho^{(j)}(\lambda)$ and $\rho_h^{(j)}(\lambda)$. 
\subsubsection{Generalized Hydrodynamics}
In the presence of an inhomogeneity (such as the kink state \eqref{initial-state}), the Bethe ansatz solution of the model \eqref{xxz-model2} breaks down due to the absence of translational invariance. Nevertheless, by considering a hydrodynamic limit for the interacting spin chain similarly to Sec.~\ref{hydro}, one can describe the asymptotic properties of the model as a collection of periodic boxes of size $\Delta x$, each containing a large number of lattice sites. In this way, the local properties of the initial non-homogeneous state can be determined in terms of a local occupation function $n^{(j)}_0(x,\lambda)$, obtained from the TBA solution of the model within the cell $\Delta x$. The large-scale dynamics of the occupation function is then established by the GHD \cite{Bertini2016,Castro-Alvaredo2016}
\be\label{GHD}
(\de_t+v^{(j)}_\text{eff}(t,x,\lambda) \ \de_x)n^{(j)}_t(x,\lambda)=0.
\ee
This set of equations has the same structure of Eq.~\eqref{Moyal} for the non-interacting model, but it is characterized by an effective velocity $v^{(j)}_\text{eff}$ dressed by the interactions, which depends self-consistently on the macrostate $n^{(j)}_t(x,\lambda)$. A formal solution of the GHD equations \eqref{GHD} is obtained with the method of characteristics, yielding
\be
n^{(j)}_t(x,\lambda)=n_0(\tilde{x}_t,\lambda)
\ee
where
\be
\tilde{x}_t=x- \int_0^t \dd t' \ v^{(j)}_\text{eff}(t',\tilde{x}_{t'},\lambda).
\ee
We mention also Ref.~\cite{Doyon-geom}, where a geometric approach for the solution of GHD equations has been derived.  In general, the solution of Eq.~\eqref{GHD} is obtained by a numerical implementation.  However, in the specific case of non-homogeneous initial macrostates that are locally described by a fully-polarized spin configuration (as it is the DW initial state under analysis), the Bethe ansatz solution of the model greatly simplifies and analytical solution of the GHD can be determined.

\subsubsection{Occupation function and effective velocity of fully-polarized states}\label{sec:analytical-BA}
As anticipated, in the specific case where the local properties of the spin chain in the cell $\Delta x$ are described by the ensemble
\be
\hat\varrho(x)=\frac{e^{2h(x)\sum_{i\in \Delta x} \ssz_i}}{{\rm tr}[e^{2h(x)\sum_{i\in \Delta x} \ssz_i}]},
\ee
where $h(x)$ is an external magnetic field, the local macrostates $n^{(j)}(x,\lambda)$ can be analytically determined, see Refs.~\cite{Takahashi-book,Collura2017,Scopa2022}. The expressions for $n^{(j)}(x,\lambda)$ are not very instructive and therefore are not reported here. The important information that  they reveal is that the occupation functions of strings $j=1,\dots, \delta-2$ are exponentially suppressed by the magnetic field while the strings $\delta-1,\delta$ are not. In the limit of strong magnetic field where
\be
\lim_{h(x)\to\pm\infty}\hat\varrho(x)=\vert\dots \uparrow\uparrow\rangle\langle\uparrow\uparrow\dots\vert \quad \text{or} \quad \vert\dots \downarrow\downarrow\rangle\langle\downarrow\downarrow\dots\vert
\ee
depending on the sign of $h(x)$, the expression for $n^{(j)}(\lambda,x)$ drammatically simplifies and reads as
\be
n^{(j)}(x,\lambda) =\begin{cases}
\delta_{j,\delta-1}+\delta_{j,\delta}, \quad \text{if $h(x)>0$};
\\[3pt]
0, \quad\text{otherwise}
\end{cases},
\ee
signaling that only the two largest strings $j=\delta-1,\delta$ are responsible for the thermodynamic properties of the fully-polarized cell $\Delta x$. It is then easy to see that the DW initial state for the interacting chain is described by the occupation functions
\be\label{DW-interacting}
n^{(j)}_0(x,\lambda) =\begin{cases}
\delta_{j,\delta-1}+\delta_{j,\delta}, \quad \text{if $x\leq 0$};
\\[3pt]
0, \quad\text{otherwise}.
\end{cases}
\ee
From Eq.~\eqref{GHD}, one can notice that the GHD evolution of the state \eqref{DW-interacting} is also determined by the sole behavior of the strings $j=\delta-1,\delta$. Furthermore, a careful inspection of the TBA equation of the model for fully-polarized states reveals that 
\be\label{eff-vel-DW}
v_\text{eff}^{(\delta-1)}=v_\text{eff}^{(\delta)}\equiv v_\delta(\lambda)=\zeta_0\sin(k_\delta(\lambda))
\ee
and the time-evolved occupation function
\be\label{n-t-DW}
n^{(\delta-1)}_t(x,\lambda)=n^{(\delta)}_t(x,\lambda)=n^{(\delta)}_0(x-v_\delta(\lambda)t,\lambda), \quad n^{(j<\delta -1)}_t(x,\lambda)=0.
\ee
We refer to Refs.~\cite{Collura2017,Collura2020,Scopa2022} for details on the derivation of these results. In Eq.~\eqref{eff-vel-DW}, we introduced  
\be
\zeta_0\equiv\sin(\gamma)/\sin(\pi/P)
\ee
and the quantity
\be
k_\delta(\lambda)\equiv s_\delta p_\delta(\lambda), 
\ee
which, up to the string sign $s_\delta$, is equal to the bare physical momentum $p_\delta(\lambda)$ of the string, see e.g. Refs.~\cite{Collura2017,Collura2020,Scopa2022} for details.\\

Eqs.~\eqref{eff-vel-DW}-\eqref{n-t-DW} allow for the derivation of analytical results for the transport properties of the interacting chain \eqref{xxz-model} during the DW melting, as discussed in the next paragraph.
\subsubsection{Analytical results}\label{sec:analytical-GHD}
As already noticed for the non-interacting case, the solution \eqref{n-t-DW} for the time-evolved occupation function is redundant due to the zero-entropy condition preserved by the GHD equations \eqref{GHD} at any time during the melting dynamics. Therefore, in analogy with the non-interacting case, we define a local Fermi contour $\Gamma_t(\Delta)$ generalizing Eq.~\eqref{Fermi-contour} as \cite{Scopa2022}
\be\label{Fermi-interacting}
\Gamma_t(\Delta)=\bigcup_x [\lambda_F^-(x,t), \lambda_F^+(x,t)]
\ee
with Fermi rapidities $\lambda_F^\pm$ at space-time position $(x,t)$ obtained by the solution of the equation of motion
\be\label{eq-of-motion-int}
x-t\ v_\delta(\lambda_F)= x- \zeta_0 t \ \sin(k_\delta(\lambda_F)) =x_0.
\ee
By noticing that the function $k_\delta(\lambda)$ is monotone in the interval $[-\pi/P,\pi/P]$ \cite{Collura2017,Collura2020,Scopa2022}, one can solve Eq.~\eqref{eq-of-motion-int} with $x_0\equiv 0$ as
\be
 k_\delta(\lambda_F^-)=\arcsin(\nicefrac{x}{\zeta_0 t}), \quad \text{if $\vert x/t\vert\leq \sin(\gamma)$}.
\ee
The other root $k_\delta(\lambda_F^+)$ comes from a Fermi rapidity $\lambda_F^+\to\infty$ initially located at $x_0<0$, corresponding to the momentum
\be
 k_\delta(\lambda_F^+)=\pi/P.
\ee
Similarly, one can show that the other string, say $\delta-1$, is related to Fermi points
\be
k_{\delta-1}(\lambda_F^\pm)=\{\pi-\arcsin(\nicefrac{x}{\zeta_0 t}), \pi-\pi/P\}.
\ee
Both  strings $j=\delta,\delta-1$ contribute with equal weight $\ell_{\delta-1}=\ell_\delta=P/2$ and, for $P=2$, they properly reproduce the situation in absence of interactions \cite{Collura2020}.
Notice that the structure of the Fermi contour $\Gamma_t$ is very similar to that of the non-interacting spin chain, although the presence of interactions lead to a shrinking of the light cone region from $-1\leq \nicefrac{x}{t}\leq 1$ at $\Delta=0$ to \cite{Collura2017,Collura2020,Scopa2022}
\be\label{eq:light-cone-shrink}
\text{light cone region}: \quad \zeta\equiv\nicefrac{x}{t}\in[-\sqrt{1-\Delta^2}, \sqrt{1+\Delta^2}].
\ee
%%%%%%%%%%%%%%%%%%%%%%%%%%%%%%%%%
\begin{figure}[t]
\centering
\includegraphics[width=\textwidth]{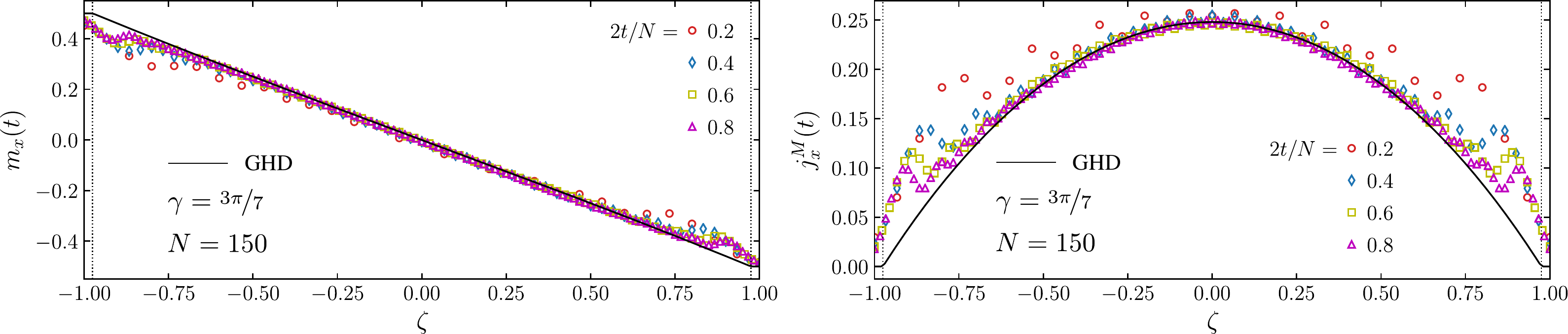}
\includegraphics[width=\textwidth]{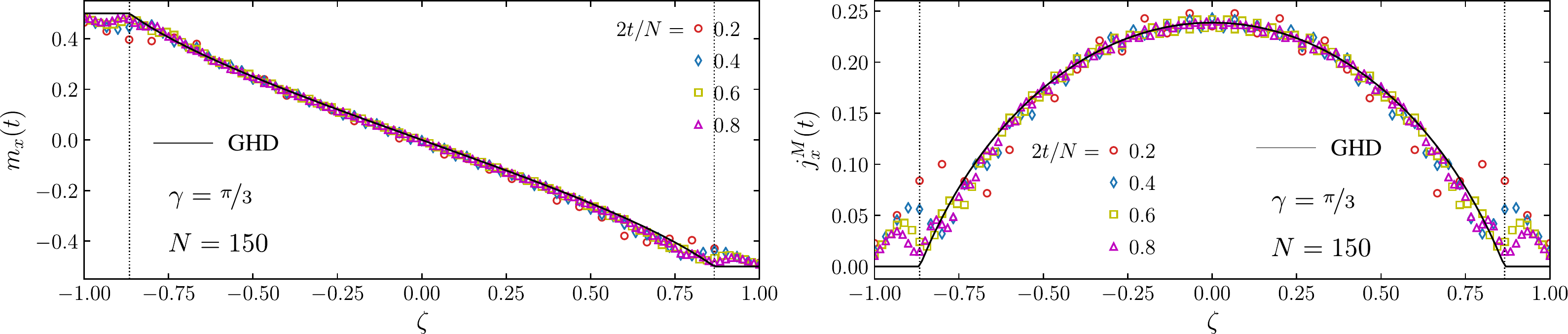}
\includegraphics[width=\textwidth]{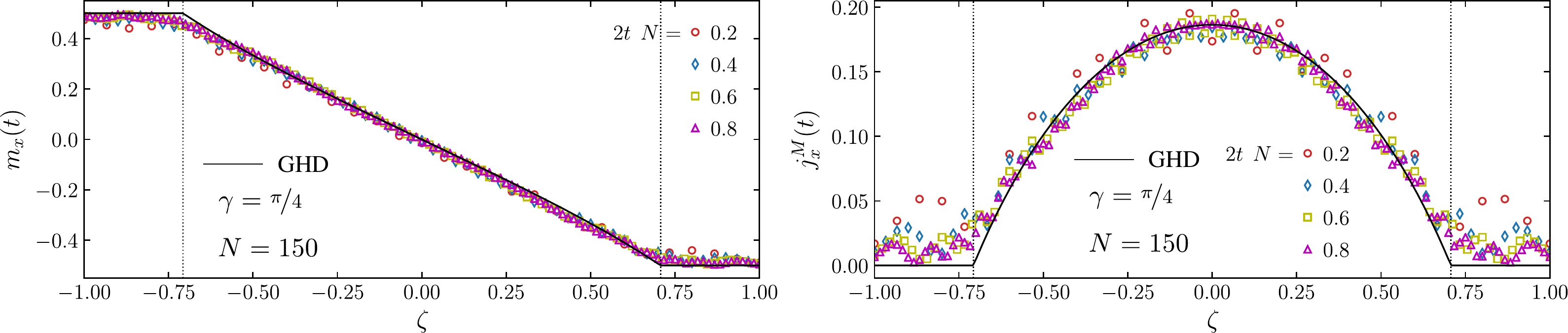}
\caption{Asymptotic profiles of magnetization (left panels) and of spin current (right panels) as function of the ratio $\zeta=\nicefrac{x}{t}$, plotted for different values of interactions. From top to bottom: $\gamma=\nicefrac{3\pi}{7},\nicefrac{\pi}{3}, \nicefrac{\pi}{4}$ corresponding to the anisotropy parameter $\Delta=0.222,0.5,0.707$. The symbols show the numerical data obtained with tensor network simulations for a spin chain of $N=150$ sites, while the full lines show the analytical GHD prediction in Eqs.~\eqref{eq:magn-GHD}-\eqref{eq:curr-GHD}. The agreement of the curves is seen extremely good, especially at large times where the convergence towards the GHD is improving. We do observe some small finite-size effects at the edges of the light cones that can be minimized by considering larger spin chains. Though, this task is found non-trivial due to the increasing of entanglement during the dynamics, which significantly slow down tensor-network based simulations, see next section. Finally, we notice that the light cone region shrinks by increasing the interaction (from top to bottom panels), see Eq.~\eqref{eq:light-cone-shrink}. In each panel, we marked the light cone positions by a dashed vertical axes. }\label{fig:profiles-dmrg}
\end{figure}
%%%%%%%%%%%%%%%%%%%%%%%%%%%%%%%%%
The charge profiles during the DW melting process follow straightforwardly. In particular the profile of magnetization for $\vert\zeta\vert\leq\sqrt{1-\Delta^2}$ is obtained as the weighted integral sum over the available quasiparticles as \cite{Collura2017,Scopa2022}
\be\begin{split}\label{eq:magn-GHD}
m_t(x)=&-\frac{1}{2} +\sum_{j=\delta-1}^\delta\ell_j \int_{-\infty}^\infty \frac{\dd \lambda}{2\pi} \ k'_j(\lambda) \ n^{(j)}_t(x,\lambda) \\[3pt]
&=-\frac{1}{2} +\frac{P}{2\pi}\int_{k_\delta(\lambda_F^-)}^{k_\delta(\lambda_F^+)} \dd k_\delta(\lambda)=
-\frac{P}{2\pi}\arcsin(\nicefrac{\zeta}{\zeta_0}).
\end{split}\ee
Outside the light cone $\zeta>\sqrt{1-\Delta^2}$ (resp.~$\zeta<-\sqrt{1-\Delta^2}$), the system keeps its initial value of magnetization $m=-1/2$ (resp.~$m=1/2$). Similarly, the spin current profile for $\vert\zeta\vert\leq \sqrt{1-\Delta^2}$ is obtained as \cite{Collura2017,Scopa2022} (see also Ref.~\cite{Pozgay2020,Pozgay2020b,Borsl2021} for a rigorous derivation of \eqref{eq:curr-GHD} in spin chain models)
\be\begin{split}\label{eq:curr-GHD}
&j^{M}_t(x)=\sum_{j=\delta-1}^\delta \ell_j\int_{-\infty}^\infty \frac{\dd \lambda}{2\pi} \ k'_j(\lambda) \  v_\text{eff}^{(j)}(t,x,\lambda) \ n^{(j)}_t(x,\lambda)\ \\[3pt]
&=\frac{P}{2\pi}\int_{k_\delta(\lambda_F^-)}^{k_\delta(\lambda_F^+)}\dd k_\delta(\lambda) v_\delta(\lambda)=
\frac{\zeta_0 P}{2\pi}\left(\sqrt{1-\nicefrac{\zeta^2}{\zeta_0^2}}-\cos(\pi/P)\right)
\end{split}\ee
and it vanishes outside the light cone. One can notice that for $Q=1$ and $P=2$ corresponding to $\Delta=0$, the results in Eqs.~\eqref{scaling-magn}-\eqref{asy-current} are recovered. In Fig.~\ref{fig:profiles-dmrg}, we show the comparison of the analytical GHD prediction in Eqs.~\eqref{eq:magn-GHD}-\eqref{eq:curr-GHD} against tensor network numerical simulations for the interacting spin chain, performed with the open-source libray iTensor \cite{iTensor}. The agreement of the curves with the numerical data is extremely good.
%
%
%
%
%
%%%%%%%%%%%%%%%%%%%%%%%%%%%%%%%%
\section{Quantum description and entanglement dynamics}\label{sec:QGHD}
The goal of this section is to complement the discussion on the transport properties of the DW melting with a study on entanglement.  As we shall review in the following paragraphs, an ab-initio characterization of entanglement with exact lattice calculations is very demanding also in absence of interaction, the case $\Delta=0$, due to the inhomogeneous and non-equilibrium character of the problem. In addition, even the hydrodynamic evolution established by the local occupation function in Eqs.~\eqref{Moyal}-\eqref{GHD} is not sufficient for the study of quantum correlations among different cells, hence of entanglement. The reason of this failure is rooted in the assumption that underlies the hydrodynamic limit considered so far. In particular, in deriving the  hydrodynamic picture, we assumed that each cell is described by a local density matrix of the form of a GGE, i.e., $\hat\varrho_t(x)\propto \exp[-\sum_{i} \beta_i(x,t) \hat{\cal Q}_i]$ in terms of some local Lagrange multipliers $\beta(x,t)$ associated with each conserved quantity, see e.g. Refs.~\cite{Castro-Alvaredo2016,ghd-notes} for more details. However, by doing this, long-range quantum coherent effects among different coarse grained points are washed out, resulting in the vanishing of equal-time correlation functions and of zero-temperature entanglement.\\

Nevertheless, a possibility to restore these missing quantum effects is given by the novel framework of Quantum Generalized Hydrodynamics (QGHD)~\cite{Ruggiero2019,Ruggiero2020,Collura2020,Scopa2021a,Scopa2022,Scopa2022b,Scopa2022c,Ruggiero2022}. According to this theory, the relevant processes at low energies are in the form of particle-hole excitations generated near the local Fermi points and can be described by a non-standard Luttinger liquid living along the evolving Fermi contour. In the following, we briefly revisit the QGHD framework and detail the solution for the DW melting problem. For the sake of clarity, we shall treat first the case $\Delta=0$ and afterwards extend the discussion on entanglement to the interacting spin chain.
\subsection{Case: $\Delta=0$}\label{sec:ent-free}
In absence of interactions, it is well known that the entanglement properties of the lattice model \eqref{xx-model} can be related to the spectrum of the two-point correlation function \eqref{eq:two-point-func}, see Ref.~ \cite{p-12,Peschel1999a,Chung2001,Peschel2003,Peschel2004,Peschel2009} for a discussion. In particular, given a bipartition of the spin chain as $A\cup B=[-\nicefrac{N}{2}+1,j]\cup[j+1,\nicefrac{N}{2}]$, one has the expression for the R\'enyi entropies
\be
S^{(\text{p})}_A=\frac{1}{1-\text{p}} \sum_{l=1}^{\nicefrac{N}{2}-1-j} \log[\Lambda^\text{p}_l+ (1-\Lambda_l)^\text{p}]
\ee
and for the entanglement entropy
\be\label{SA-num}
S_A=-\sum_{l=1}^{\nicefrac{N}{2}-1-j} [\Lambda_l\log\Lambda_l + (1-\Lambda_l)\log(1-\Lambda_l)].
\ee
Here, $\Lambda_l$ are the eigenvalues of the two-point correlation function \eqref{eq:two-point-func} restricted to the subsystem $A$, i.e., $G_A\equiv[G_{n,m}]_{n,m\in A}$. However, such a direct approach is very demanding and often not possible in inhomogeneous quench settings as the one that we are considering. In some special cases, some results have been obtained following this strategy (see e.g. Refs.~\cite{Eisler2008,Eisler2014,Eisler2016,Eisler2017,Eisler2021}) but these are based on non-trivial lattice calculations and therefore will not be discussed in this context.\\
\subsubsection{Re-quantization of the Fermi contour}
In this short review, we rather proceed by considering a re-quantization of the Euler hydrodynamics of Sec.~\ref{sec:transport} including linear quantum fluctuations at the edges of the local Fermi points as
\be\label{density-fluc}
k_F(x,t)\ \mapsto \ k_F(x,t)+\delta\hat{k}_t(x).
\ee
and define the excess density simply as $\delta\hat{n}_t(x)=\delta\hat{k}_t(x)/\pi$. Using standard bosonization arguments \cite{Giamarchi2007,Cazalilla2004,Cazalilla2011}, one can then relate the fluctuating field $\delta\hat{n}_t(x)$ to the vertex operators of an effective field theory arising at large scales and at low energies. To this end, we express $\delta\hat{n}_t(x)$ as 
\be
\delta\hat{n}_t(x)=\frac{1}{\pi} \de_x \hat\phi_t(x),
\ee
where $\hat{\phi}_t(x)$ is a height field encoding the long-range density fluctuations along the one-dimensional spin chain, see e.g. Ref.~\cite{Brun2017,Brun2018,Bastianello2020,Scopa2020}. From Eq.~\eqref{density-fluc}, one can obtain the leading order term of the Haldane harmonic-fluid expansion for the time-dependent lattice fermions as \cite{Ruggiero2019,Scopa2021a,Ruggiero2022}
\be
\hat{c}^\dagger_x(t)\sim C(x,t) e^{-\I \varphi(x,t)} \text{\bf:}\ e^{\I \hat\theta_t(x)}\ \textbf{:} +\text{h.o.c.}
\ee
with dimensionful non-universal amplitude $C(x,t)$ and semi-classical phase $\varphi(x,t)$ that are unimportant for our scopes, see e.g. Refs.~\cite{Dubail2017,Brun2018,Ruggiero2019,Scopa2020} for a discussion. Here, \textbf{:}$\ \cdot\ $\textbf{:} denotes the normal ordering of fields and $\hat\theta_t(x)$ is a phase-fluctuating field satisfying
\be
[\hat\theta_t(x),\hat\phi_t(x')]=\frac{\I\pi}{2} {\rm sgn}(x-x').
\ee
Higher order terms are obtained from vertices with higher scaling dimension but their contribution is negligible in the low-energy regime. It is then customary to express the fields $\hat\phi_t(x)$ and $\hat\theta_t(x)$ in terms of two chiral bosons $\hat\phi^{(\pm)}$ \cite{Cazalilla2004}
\be\label{Haldane}
\hat\phi_t(x)\equiv \frac{1}{2 \sqrt{K_t(x)}}[\hat\phi^{(+)}_t(x)+\hat\phi^{(-)}_t(x)]; \quad \hat\theta_t(x)\equiv\frac{\sqrt{K_t(x)}}{2}[\hat\phi^{(+)}_t(x)-\hat\phi^{(-)}_t(x)],
\ee
describing left- and right- moving propagating sound waves along the Fermi contour. The parameter $K_t(x)$ is usually called {\it Luttinger parameter} and is related to the local compressibility of the off-equilibrium quantum fluid. In absence of interactions,
\be\label{K=1}
\text{free fermionic limit:} \quad K_t(x)\equiv 1
\ee
irrespective on the space-time position. At this point, by plugging the expansion \eqref{Haldane} in \eqref{xx-model} and retaining only quadratic terms, one arrives to the Luttinger liquid Hamiltonian
\be\label{H-ll}
\Ha_\text{LL}=\frac{1}{2\pi}\int \dd x \ v_t(x) \left[(\de_x\hat\theta_t(x))^2 + (\de_x\hat\phi_t(x))^2\right]
\ee
where the sound velocity $v_t(x)\equiv\sin k_F(x,t)$. In terms of the chiral fields,
\be
\Ha_\text{LL}=\frac{1}{2\pi}\int \dd x \left[\sin k_F^+(x,t) (\de_x \hat\phi^{(a[+])}_t(x))^2 + \sin k_F^-(x,t) (\de_x \hat\phi^{(a[-])}_t(x))^2\right] 
\ee
with $a[\pm]\equiv a(k_F^\pm(x,t))$ and $a(k)=\mp$ if $\text{sign}(k)\lessgtr 0$. Finally, from the expression \eqref{H-ll} for the Luttinger liquid Hamiltonian, one obtains the action \cite{Allegra2016,Dubail2017,Brun2018,Scopa2020,Bastianello2020,Collura2020}
\be\label{action}
\mathscr{S}=\int \frac{\sqrt{\det(g)} \ \dd t \ \dd x}{2\pi K_t(x)} g^{ab} \de_a\hat\phi_t(x) \de_b\hat\phi_t(x) 
\ee
where the indices $a,b=x,t$ and we restored the dependence on $K_t(x)\equiv 1$ for future convenience. The action in Eq.~\eqref{action} describes a free massless compact boson with space-time dependent coupling $K_t(x)$ and time-dependent non-flat 2-dimensional metric tensor $g_{ab}$, whose line element reads as 
\be\label{metric}
\dd s^2= \left[ v_t(x) \ \dd t - \dd x\right]^2.
\ee
It is easy to show that this metric can be mapped to a flat one with a simple change of coordinates where $\dd x\mapsto \dd\tilde{x}=\dd x/v_t(x)$, see e.g. Ref.~\cite{Allegra2016,Dubail2017,Brun2018,Scopa2020,Bastianello2020,Collura2020,Scopa2021a} for details. Hence, the action \eqref{action} displays conformal invariance in those settings where $K_t(x)=\text{const}$, as it is the case for a non-interacting spin chain (cf.~Eq.~\eqref{K=1}). In the following, we shall make use of tools stemming from conformal field theory (CFT) to establish the universal behavior of the entanglement entropy during the DW melting problem.
\subsubsection{Universal behavior of entanglement}
The universal behavior of R\'enyi entropy for a bipartition $A\cup B\equiv[-\infty, x_0]\cup(x_0,+\infty]$ with cutting point in a real-space position $x_0$ can be established using the known relation \cite{Calabrese2004,Cardy2008,Calabrese2009}
\be\label{cft-renyi}
\tilde{S}^{(\text{p})}_{x_0}=\frac{1}{1-\text{p}}\log[{\rm tr}[\hat\rho_{x_0}]^{\text{p}}]=\frac{1}{1-\text{p}} \log \langle \hat{\cal T}^{(\text{p})}_t(x_0)\rangle,
\ee
obtained from the path-integral representation of the operator $[\hat\rho_{x_0}]^{\text{p}}$ on a $\text{p}$-sheeted Riemann surface. Here, $\hat{\cal T}_t^{(\text{p})}(x_0)$ is a boundary field  known as {\it twist field} connecting local operators among the $\text{p}$-copies of the model across the branch-cut at $x_0$, see e.g. Refs.~\cite{Cardy2008,Calabrese2009} for a discussion. Crucially, in our model, the twist field is a primary operator of the CFT along the Fermi contour with scaling dimension
\be
d_\text{p}=\frac{1}{12}(\text{p}-\nicefrac{1}{\text{p}})
\ee
and it allows for a chiral decomposition
\be
\hat{\cal T}^{(\text{p})}_t(x_0) = \hat\tau^{(\text{p})}_t(x_0,+) \otimes \hat\tau^{(\text{p})}_t(x_0,-)
\ee
in terms of the chiral components $\hat\tau^{(\text{p})}_t(x_0,\pm)$, each with scaling dimension $d_\text{p}/2$. Notice that these objects are highly non-trivial if compared to their equilibrium counterparts, due to their time dependence and to the non-homogeneous background over which expectation values are evaluated. However, a natural description for these boundary fields is attained by moving from space-time coordinates $(x,t)$ to a coordinate $s\equiv s(x,t)$ along the Fermi contour $\Gamma_t$ at time $t$. By doing this, Eq.~\eqref{cft-renyi} becomes
\be
\tilde{S}^{(\text{p})}_{x_0}=\frac{1}{1-\text{p}} \log\left[\left\vert\frac{\dd s}{\dd x}\right\vert_{s\equiv s_+}^{d_\text{p}/2}\left\vert\frac{\dd s}{\dd x}\right\vert_{s\equiv s_-}^{d_\text{p}/2} \langle \hat\tau^{(\text{p})}(s_+)\hat\tau^{\text{(p)}}(s_-)\rangle\right],
\ee
where $s_\pm$ are the boundary points of the subsytem $A$ along $\Gamma_t$, see the discussion below. Notice that in terms of the coordinate $s$, sound waves can be described by a unique component circulating along the contour, see e.g. Refs.~\cite{Scopa2021a,Ruggiero2022}. The expectation value in the last equation is then obtained by standard boundary CFT and reads as
\be\label{eq-renyi-s}
\langle \hat\tau^{(\text{p})}(s_+)\hat\tau^{\text{(p)}}(s_-)\rangle= \left\vert 2\sin\left(\frac{s_+-s_-}{2}\right)\right\vert^{-d_\text{p}}.
\ee
The coordinate $s(x,t)$ is an isothermal coordinate for the space-time dependent metric \eqref{metric}. Its explicit expression is generically out-of-reach, excluding outstanding cases such as Ref.~\cite{Ruggiero2019} and  Refs.~\cite{Allegra2016}. For generic interacting problems, one typically finds an isothermal coordinate $s(x,0)$ for the initial inhomogeneous configuration as done in Refs.~\cite{Brun2017,Brun2018,Scopa2020,Bastianello2020} and evolves the initial correlations in time according to the equation derived in Ref.~\cite{Ruggiero2020}. For free systems, the isothermal coordinate $s(x,0)$ is a comoving coordinate since quantum fluctuations in the initial state do not evolve and are only transported along the contour $\Gamma_t$ during the dynamics,  see Refs.~\cite{Dubail2017,Scopa2021a,Scopa2022,Scopa2022b,Scopa2022c} for examples.\\
%%%%%%%%%%%%%%%%%%%%%%%%%%
\begin{figure}[t]
\centering
\includegraphics[width=.9\textwidth]{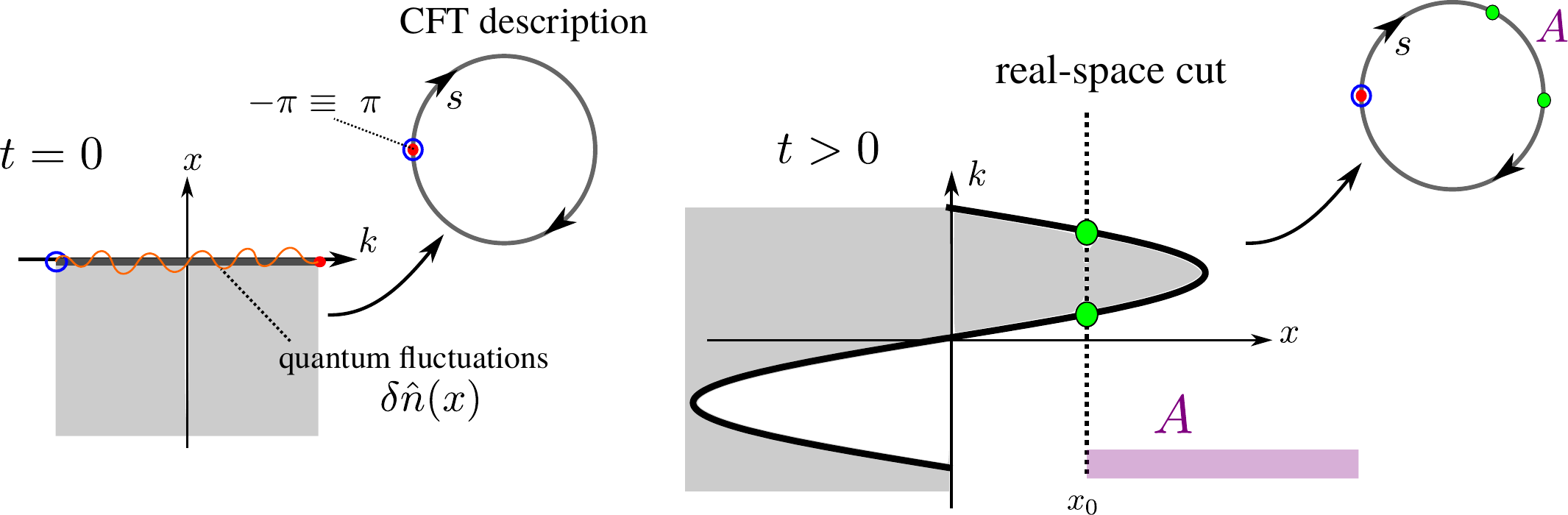}
\caption{Illustration of the QGHD framework for the DW melting problem. Left~--~At $t=0$, we include linear quantum fluctuations $\delta\hat{n}(x)$ on top of the initial Fermi contour at $x=0$ and we map momentum-space correlations into a CFT that lives along a unit circle of length $2\pi$. Right~--~ At $t>0$, through the ballistic propagation of particles from the junction, the system develops real-space entanglement for any bipartition with cutting point $x_0$ inside the light cone. The amount of entanglement in real space is then established by the local Fermi points at $(x,t)$ (green circles) and simply translates into the entanglement of the interval $[s_-,s_+]$ for the CFT along the Fermi contour, thanks to the non-interacting nature of the problem.}\label{fig:ent-ill}
\end{figure}
%%%%%%%%%%%%%%%%%%%%%%%%%%

The DW melting problem falls in the simplest category. An isothermal coordinate for this quench is simply given by \cite{Allegra2016,Dubail2017}
\be\label{iso-DW}
s(x,t)\equiv k_F(\zeta)=\arcsin(\zeta), \qquad -1< \zeta< 1.
\ee
It follows that the boundary points $s_\pm$ for a real-space cut at position $x_0$ are given by
\be
s_\pm=\{ \pi-\arcsin(\nicefrac{x_0}{t}); \ \arcsin(\nicefrac{x_0}{t})\}
\ee
and, by elementary algebra, one obtains directly from Eq.~\eqref{eq-renyi-s} the result \cite{Dubail2017}
\be\label{result-univ-free}
\tilde{S}^{(\text{p})}_{x_0}=\frac{\text{p}+1}{12\text{p}}\log\left[2t\left\vert 1-\nicefrac{x_0}{t}\right\vert \right].
\ee
Despite its simplicity, Eq.~\eqref{iso-DW} allows for a non-trivial interpretation that we wish to comment. In fact, the change of coordinates $(x,t)\mapsto s\equiv k_F(\zeta)$  implements a mapping between real-space correlations and momentum-space correlations of the spin chain during the melting dynamics. At $t=0$, we prepare our system in a product state, i.e., with exact zero entanglement for any cutting position $x_0$. However, the same does not hold true if one considers the entanglement of a bipartition in momentum space, for which standard boundary CFT can be applied. During the time evolution, momentum-space entanglement is  gradually transported to real-space through the ballistic propagation of particles, resulting in the logarithmic growth of Eq.~\eqref{result-univ-free}. In this sense, the isothermal coordinate $s(x,t)$ may be viewed as a bridge between real and momentum space, and therefore it establishes the amount of spatial correlations that are generated via propagation at a given space-time point $(x_0,t)$. We illustrate this procedure in Fig.~\ref{fig:ent-ill}.
\subsubsection{Short distance regularization and asymptotic results}
At this point, we recall that Eq.~\eqref{result-univ-free} provides only the universal contribution to the R\'enyi entropies and it has to be complemented with a non-universal regularization at short distances \cite{Dubail2017}. The latter, for homogeneous equilibrium models, typically introduces an additive constant and therefore is often neglected or estimated as fitting parameter. Though, in out-of-equilibrium and non-homogeneous situations, its effect is more pronounced and significantly modify the space-time dependence of R\'enyi entropies in Eq.~\eqref{result-univ-free}. Hence, by dimensional analysis, we can write the regularized R\'enyi entropy as
\be\label{tot-renyi}
S^{(\text{p})}_{x_0}=\frac{1}{1-\text{p}} \log \langle (\varepsilon_{x_0}(t))^{d_\text{p}}\ \hat{\cal T}^{(\text{p})}_{x_0}(t)\rangle =\tilde{S}^{(\text{p})}- \frac{\text{p}+1}{12\text{p}} \log[ \varepsilon_{x_0}(t)]
\ee
where $\varepsilon_{x_0}(t)$ is a local UV cutoff. On physical grounds, one expects that the short-distance regularization is set by the microscopic scale of the model within the cell $x_0$, which is the inverse local density, and reads as
\be\label{guess}
\varepsilon_{x_0}(t)\propto (\nicefrac{1}{2}+m_{x_0}(t))^{-1}.
\ee
From exact Fisher-Hartwig calculations  made for the microscopic XX model \cite{Calabrese2010,Jin2004}, one  can refine the ansatz in Eq.~\eqref{guess} and obtains the correct value of cutoff as \cite{Dubail2017,Scopa2021a,Scopa2022,Scopa2022b,Scopa2022c,Ruggiero2022}
\be\label{cutoff}
\varepsilon_{x_0}(t)= \frac{C}{\cos(\pi m_{x_0}(t))},
\ee
with $C$ a known non-universal amplitude, see Refs.~\cite{Jin2004,Calabrese2010} for its expression. Using Eq.~\eqref{cutoff} in Eq.~\eqref{tot-renyi}, after simple algebra one obtains the final result
\be\label{ent-DW-free}
S^{(\text{p})}_{x_0}=\frac{\text{p}+1}{12\text{p}}\log[t(1-\nicefrac{x_0^2}{t^2})^{3/2}] + \kappa_\text{p},
\ee
where the additive constant $\kappa_\text{p}=-\nicefrac{(\text{p}+1)}{12\text{p}}\log[C/2]$, for instance $\kappa_1=0.4785$ \cite{Jin2004}. %\textcolor{red}{Notice that Eq.~\eqref{ent-DW-free} correctly predicts zero entanglement at $t=0$ {\tt "this is not clear to me how it can be so"}}and, for $t>0$, a non-vanishing entanglement in the light cone region $-t\leq x_0\leq t$, in agreement with our intuitions. 
By setting $x_0=0$ and $\text{p}=1$ in Eq.~\eqref{ent-DW-free}, we see that the half-system entanglement entropy displays a logarithmic growth in time as \cite{Dubail2017,Scopa2022,Scopa2022b}
\be\label{half-sys-free}
S_{0}(t)= \frac{1}{6}\log(t) + \kappa_1.
\ee
This results can be alternatively viewed as a manifestation of Calabrese-Cardy formula \cite{Calabrese2004} (see also Ref.~\cite{Holzhey-Larsen-Wilczek} by Holzhey-Larsen-Wilczek), since the size of the correlated region is growing linearly with time in our problem. \\

In Fig.~\ref{fig:ent-free}, we show the entanglement profiles \eqref{ent-DW-free} and the half-system entanglement entropy \eqref{half-sys-free} against exact numerical data and we observe a perfect matching of the curves with numerics.\\

%%%%%%%%%%%%%%%%%%%%%%%%%%
\begin{figure}[t]
\centering
\includegraphics[width=\textwidth]{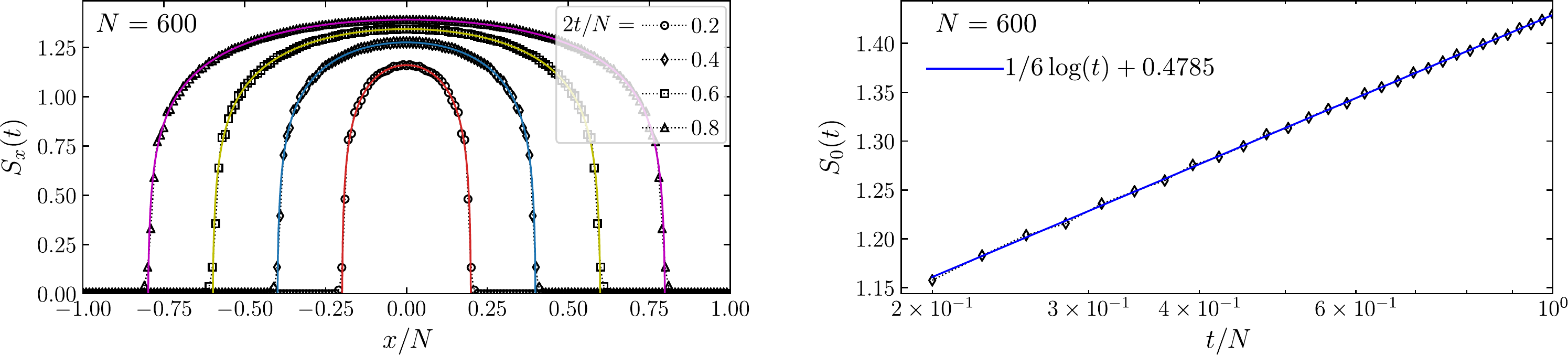}
\caption{Left~--~Entanglement entropy for the DW melting of a non-interacting spin chain, plotted as function of the cutting point $x$ and for different times. Symbols show the numerical data obtained from Eq.~\eqref{SA-num} after the exact diagonalization of $G_A$ while the full lines show the predictions by QGHD of Eq.~\eqref{ent-DW-free}. Right~--~Half-system entanglement growth during the DW melting as function of time, plotted in semi-logarithmic scale. }\label{fig:ent-free}
\end{figure}
%%%%%%%%%%%%%%%%%%%%%%%%%%
\subsection{Case: $-1<\Delta<1$}
As anticipated, in generic interacting integrable models the Luttinger parameter $K_t(x)$ is non-constant and non-homogeneous and, therefore, the conformal invariance of the action \eqref{action} is generically lost. Its value can be determined (as at equilibrium) from the value of the dressed magnetization evaluated at any of the local Fermi rapidities $\lambda_F^\pm(x,t)$ in Eq.~\eqref{Fermi-interacting}, see Ref.~\cite{Collura2020} for a discussion and e.g. Ref.~\cite{Korepin2010} for details on the calculation. \\ 

However, the very peculiar features about the Bethe ansatz of the DW state (cf. Refs.~\cite{Collura2017,Scopa2022,Takahashi-book} and Sec.~\ref{sec:analytical-BA}) that led to the very outstanding possibility of finding an analytical solution for the GHD equation of the model (cf.~Refs.\cite{Collura2017,Collura2020,Scopa2022} and Sec.~\ref{sec:analytical-GHD}), permit the analytical calculation of the Luttinger parameter, which for both the strings $j=\delta-1, \delta$ equals to
\be
K_t(x)\equiv K= \frac{P^2}{4}.
\ee
Remarkably, the Luttinger parameter for the DW melting problem is constant and depends only on the interaction parameter (actually only on the denominator, recall that $\gamma=\arccos(\Delta)=\pi Q/P$), displaying a peculiar fractal dependence \cite{Collura2020}.

Consequently, conformal invariance in Eq.~\eqref{action} is not broken even at finite interactions and one can obtain the R\'enyi entropy of a bipartition with cutting point $x_0$ employing the same techniques of Sec.~\ref{sec:ent-free}. In particular, one can still write the relation
\be
S_{x_0}^{(\text{p})}(\Delta)=\frac{1}{1-\text{p}}\log\langle \varepsilon_{x_0}(t,\Delta)^{d_\text{p}} \hat{\cal T}_{x_0}^{(\text{p})}(t)\rangle
\ee
and, by scaling arguments, one obtains \cite{Collura2020}
\be\label{HS-ent-int}
S_{x_0}^{(\text{p})}(\Delta)=\frac{\text{p}+1}{12\text{p}}\log(t) + f_\text{p}(\nicefrac{x}{\zeta_0 t}) + \kappa_\text{p}(\Delta)
\ee
where $f_\text{p}(\zeta/\zeta_0)$ is a scaling function that has been conjectured in Ref.~\cite{Collura2020} and $\kappa_\text{p}(\Delta)$ is a non-universal additive constant that we are currently unable to analytically determine. From the numerical analysis of Fig.~\ref{fig:ent-int}, we observe that the qualitative behavior of the entanglement for the interacting chain $\vert \Delta\vert<1$ is not limited to the logarithmic half-system growth (cf.~Eq.~\eqref{HS-ent-int} and \eqref{half-sys-free}) but instead applies to the whole entanglement profiles, modulo a rescaling by the interactions of the light cone, as pointed out in Eq.~\eqref{eq:light-cone-shrink}.
%%%%%%%%%%%%%%%%%%%%%%%%%%
\begin{figure}[t]
\centering
\includegraphics[width=\textwidth]{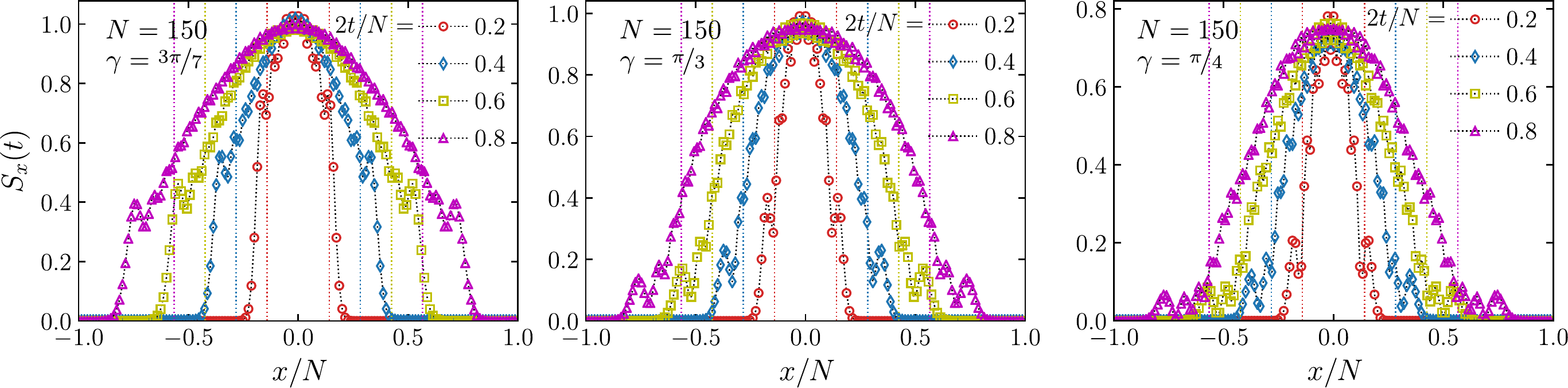}
\caption{Entanglement profiles during the DW melting of the interacting spin chain, plotted at different times and for different values of interaction (from left to right: $\gamma=\nicefrac{3\pi}{7},\nicefrac{\pi}{3},\nicefrac{\pi}{4}$ corresponding to $\Delta=0.222,0.5,0.707$). The data is obtained with tensor network simulations. Dashed vertical axes mark the position of the light cone at each time, according to Eq.~\eqref{eq:light-cone-shrink}.}\label{fig:ent-int}
\end{figure}
\section{Summary and concluding remarks}\label{sec:conclusion}
In this short review, we revisited the out-of-equilibrium physics arising during the unitary evolution of a one-dimensional spin chain prepared in a domain wall configuration. For this problem, we focused on transport properties (i.e., characterization of charge and current profiles) and on the study of entanglement entropy.\\ In Sec.~\ref{sec:transport}, the first aspect has been first addressed for a noninteracting spin chain by exact lattice methods, from which a hydrodynamic approach has been later developed. We then extended such a hydrodynamic approach to the interacting case by means of Bethe ansatz techniques and through the use of generalized hydrodynamics, both of which are briefly commented on in the text. We derived exact results for charge and current profiles and showed how these are found in perfect agreement with the numerical data, obtained by exact methods or by the use of tensor network-based numerical simulations.\\
In the second part of the review (Sec.~\ref{sec:QGHD}), we revisited the study of entanglement dynamics by means of quantum generalized hydrodynamics, which consists of the use of CFT tools arising from the requantization of the hydrodynamic background through a Luttinger liquid. Analytical results for R\'enyi entropy profiles at different times have been derived and compared with numerical data.\\

As commented in the introduction, the domain wall melting problem has been the subject of numerous studies e.g.~\cite{Antal1999,Antal2008,Karevski2002,Platini2007,Rigol2004,Platini2005,Dubail2017,Alba2014,Eisler2008,Eisler2014,Scopa2021a,Scopa2022,Scopa2022b,Scopa2022c,Rottoli2022,Allegra2016},
 from which numerous analytical results have been collected over years. Besides being one of the very rare cases of non-trivial systems where such solutions can be derived, the study of this setting is also extremely useful as  basis for investigating the dynamics of other quench problems with generic integrable models. Hence, the purpose of this review is twofold: first, to concisely provide a summary of available results for domain wall melting, and second, to give an overview accessible to a broad community of the hydrodynamic approach to integrable systems via a thorough study of a prototypical model, which has recently attracted a great deal of attention.

\subsection*{Data Availability Statement}
No Data associated in the manuscript.

\subsection*{Acknowledgments}
The authors would like to convey their best wishes to their colleague and friend Malte Henkel for his 60th birthday.\\

 SS acknowledges support from ERC under Consolidator Grant No.~771536 (NEMO). 
 DK acknowledges support from the French ANR funding UNIOPEN
(Grant No.~ANR-22-CE30-0004-01).
 SS is very thankful to the LPCT of Nancy for the invitation to the SPLDS22 conference where this work has been initiated. The authors acknowledge J. Dubail, F. Ares, P. Ruggiero, S. Wald, D. Horv\'ath, F. Rottoli, L. Capizzi and  P. Calabrese for useful discussions on QGHD and for collaborations on closely related topics.

\input{bibliography}
\end{document}

%% file: new_commands.tex
\newcommand{\be}{\begin{equation}}
\newcommand{\ee}{\end{equation}}

\newcommand{\beA}{\begin{align}}
\newcommand{\eeA}{\end{align}}

\newcommand{\bV}{\begin{pmatrix}}
\newcommand{\eV}{\end{pmatrix}}

\newcommand{\dd}{\mathrm{d}}
\newcommand{\de}{\partial}

\newcommand{\Ha}{\hat{H}} % Notation for the Hamiltonian

\newcommand{\ssz}{\hat{\sigma}^\text{z}} % Notation for spin-chain operators
\newcommand{\ssx}{\hat{\sigma}^\text{x}}
\newcommand{\ssy}{\hat{\sigma}^\text{y}}
\newcommand{\ssp}{\hat{\sigma}^+}
\newcommand{\ssm}{\hat{\sigma}^-}

%% file: main.bbl
\begin{thebibliography}{100}

\bibitem{Rigol2007} M. Rigol, V. Dunjko, V. Yurovsky and M. Olshanii,
Relaxation in a Completely Integrable Many-Body Quantum System: An Ab Initio Study of the Dynamics of the Highly Excited States of 1D Lattice Hard-Core Bosons,
\href{https://journals.aps.org/prl/abstract/10.1103/PhysRevLett.98.050405}{Phys. Rev. Lett. {\bf98}, 050405 (2007)}.

\bibitem{Bertini2016} B. Bertini, M. Collura, J. De Nardis and M. Fagotti,
 Transport in out-of-equilibrium XXZ chains: Exact profiles of charges and currents, 
\href{https://journals.aps.org/prl/abstract/10.1103/PhysRevLett.117.207201}{Phys. Rev. Lett. {\bf117}, 207201(2016)}.

\bibitem{Castro-Alvaredo2016} O. A. Castro-Alvaredo, B. Doyon and T. Yoshimura,
Emergent hydrodynamics in integrable quantum systems out of equilibrium,
\href{https://journals.aps.org/prx/abstract/10.1103/PhysRevX.6.041065}{Phys. Rev. X {\bf6}, 041065 (2016)}

\bibitem{ghd-rev} V. Alba, B. Bertini, M. Fagotti, L. Piroli and P. Ruggiero,
{ Generalized-hydrodynamic approach to inhomogeneous quenches: correlations, entanglement and quantum effects},
\href{https://iopscience.iop.org/article/10.1088/1742-5468/ac257d/meta}{J. Stat. Mech. (2021) 114004}.

\bibitem{ghd-notes} B. Doyon,
{ Lecture notes on Generalised Hydrodynamics},
\href{https://doi.org/10.21468/SciPostPhysLectNotes.18}{SciPost Phys. Lect. Notes {\bf18} (2020)}

\bibitem{Essler-ghd-rev} F.~H.~L. Essler,
{ A short introduction to Generalized Hydrodynamics}
\href{https://doi.org/10.1016/j.physa.2022.127572}{Physica A 2022, 127572}.

\bibitem{DeNardis-rev} J. De Nardis, B. Doyon, M. Medenjak and M. Panfil,
Correlation functions and transport coefficients in generalised hydrodynamics
\href{https://iopscience.iop.org/article/10.1088/1742-5468/ac3658/meta}{J. Stat. Mech. (2022) 014002}.

\bibitem{Collura2017} M. Collura, A. De Luca and J. Viti,
Analytic solution of the Domain Wall non-equilibrium stationary state,
\href{https://doi.org/10.1103/PhysRevB.97.081111}{Phys. Rev. B {\bf97}, 081111 (2018)}.

\bibitem{DeLuca2017} A. De Luca, M. Collura and J. De Nardis,
 Nonequilibrium spin transport in integrable spin chains: Persistent currents and emergence of magnetic domains, 
\href{https://journals.aps.org/prb/abstract/10.1103/PhysRevB.96.020403}{Phys. Rev. B {\bf 96}, 020403 (2017)}

\bibitem{Bastianello2019} A. Bastianello, V. Alba, and J. S. Caux
Generalized Hydrodynamics with Space- Time Inhomogeneous Interactions, 
\href{https://journals.aps.org/prl/abstract/10.1103/PhysRevLett.123.130602}{Phys. Rev. Lett. {\bf123}, 130602 (2019)}

\bibitem{Bulchandani2017} V. B. Bulchandani, R. Vasseur, C. Karrasch and J. E. Moore,
 Solvable Hydrodynamics of Quantum Integrable Systems, 
\href{https://journals.aps.org/prl/abstract/10.1103/PhysRevLett.119.220604}{Phys. Rev. Lett. {\bf 119}, 220604 (2017)}

\bibitem{Bulchandani2018} V. B. Bulchandani, R. Vasseur, C. Karrasch, and J. E. Moore,
 Bethe-Boltzmann hydrodynamics and spin transport in the XXZ chain, 
\href{https://journals.aps.org/prb/abstract/10.1103/PhysRevB.97.045407}{Phys. Rev. B {\bf97}, 045407 (2018)}

\bibitem{Doyon2017} B. Doyon, J. Dubail, R. Konik and T. Yoshimura,
 Large-Scale Description of Interacting One-Dimensional Bose Gases: Generalized Hydrodynamics Supersedes Conventional Hydrodynamics,
\href{https://journals.aps.org/prl/abstract/10.1103/PhysRevLett.119.195301}{Phys. Rev. Lett. {\bf119}, 195301 (2017)}

\bibitem{Doyon2018} B. Doyon, T. Yoshimura, and J. S. Caux,
 Soliton Gases and Generalized Hydrodynamics,
\href{https://journals.aps.org/prl/abstract/10.1103/PhysRevLett.120.045301}{Phys. Rev. Lett. {\bf120}, 045301 (2018)}

\bibitem{Piroli2017} L. Piroli, J. De Nardis, M. Collura, B. Bertini and M. Fagotti,
 Transport in out-of-equilibrium XXZ chains: Nonballistic behavior and correlation functions,
\href{https://journals.aps.org/prb/abstract/10.1103/PhysRevB.96.115124}{Phys. Rev. B {\bf 96}, 115124 (2017)}

\bibitem{Caux2019} J.S.~Caux, B. Doyon, J. Dubail, R. Konik, T. Yoshimura,
Hydrodynamics of the interacting Bose gas in the Quantum Newton Cradle setup,
\href{https://scipost.org/SciPostPhys.6.6.070}{SciPost Phys. {\bf6}, 070 (2019)}.

\bibitem{Scopa2022} S. Scopa, P. Calabrese and J. Dubail,
Exact hydrodynamic solution of a double domain wall melting in the spin-1/2 XXZ model,
\href{https://doi.org/10.21468/SciPostPhys.12.6.207}{SciPost Phys. {\bf12}, 207 (2022)}.

\bibitem{Schemmer2019} M. Schemmer, I. Bouchoule, B. Doyon, and J. Dubail,
 Generalized Hydrodynamics on an Atom Chip,
\href{https://journals.aps.org/prl/abstract/10.1103/PhysRevLett.122.090601}{Phys. Rev. Lett.{\bf 122}, 090601 (2019)}

\bibitem{Malvania2020} N. Malvania, Y. Zhang, Y. Le, J. Dubail, M. Rigol, and D. S. Weiss,
Generalized hydrodynamics in strongly interacting 1D Bose gases,
 \href{https://www.science.org/doi/abs/10.1126/science.abf0147}{Science {\bf 373}, 6559 (2021)}

\bibitem{ghd-JB} I. Bouchoule and J. Dubail,
{ Generalized hydrodynamics in the one-dimensional Bose gas: theory and experiments},
\href{https://doi.org/10.1088/1742-5468/ac3659}{J. Stat. Mech. (2022) 014003}.

\bibitem{Bertini2021} B. Bertini, F. Heidrich-Meisner, C. Karrasch, T. Prosen, R. Steinigeweg and M. Znidaric,
Finite-temperature transport in one-dimensional quantum lattice models,
\href{https://journals.aps.org/rmp/abstract/10.1103/RevModPhys.93.025003}{Rev. Mod. Phys. {\bf93}, 025003 (2021)}.

\bibitem{Nozawa2020} Y. Nozawa and H. Tsunetsugu,
{Generalized hydrodynamic approach to charge and energy currents in the one-dimensional Hubbard model},
\href{https://doi.org/10.1103/PhysRevB.101.035121}{Phys. Rev. B {\bf101}, 035121 (2020)}.

\bibitem{Nozawa2021} Y. Nozawa and H. Tsunetsugu,
{Generalized hydrodynamics study of the one-dimensional Hubbard model: Stationary clogging and proportionality of spin, charge, and energy currents},
\href{https://doi.org/10.1103/PhysRevB.103.035130}{Phys. Rev. B {\bf103}, 035130 (2021)}.

\bibitem{Mestyan2019} M. Mesty{\'{a}}n, B. Bertini, L. Piroli and P. Calabrese, 
{Spin-charge separation effects in the low-temperature transport of one-dimensional Fermi gases},
\href{https://doi.org/10.1103/PhysRevB.99.014305}{Phys. Rev. B {\bf 99},014305 (2019)}.

\bibitem{Scopa2021} S. Scopa, P. Calabrese and L. Piroli,
{Real-time spin-charge separation in one-dimensional Fermi gases from generalized hydrodynamics},
\href{https://doi.org/10.1103/PhysRevB.104.115423}{Phys. Rev. B {\bf104}, 115423 (2021)}.

\bibitem{Moller-2component} F. M{\o}ller, C. Li, I. Mazets, H.~-P. Stimming, T. Zhou, Z. Zhu, X. Chen, and J. Schmiedmayer,
{Extension of the Generalized Hydrodynamics to the Dimensional Crossover Regime},
\href{https://journals.aps.org/prl/abstract/10.1103/PhysRevLett.126.090602}{Phys. Rev. Lett. {\bf126},090602 (2021)}.

\bibitem{Scopa2022d} S. Scopa, P. Calabrese and L. Piroli,
{Generalized hydrodynamics of the repulsive spin-1/2 Fermi gas},
\href{https://journals.aps.org/prb/pdf/10.1103/PhysRevB.106.134314}{Phys. Rev. B {\bf106}, 134314 (2022)}.

\bibitem{DeNardis2018} J. De Nardis, D. Bernard and B. Doyon,
{Hydrodynamic Diffusion in Integrable Systems},
\href{https://journals.aps.org/prl/abstract/10.1103/PhysRevLett.121.160603}{Phys. Rev. Lett. {\bf 121}, 160603 (2018)}.

\bibitem{DeNardis2019} J. De Nardis, D. Bernard and B. Doyon, 
{Diffusion in generalized hydrodynamics and quasiparticle scattering},
\href{https://doi.org/10.21468/SciPostPhys.6.4.049}{SciPost Phys. {\bf6}, 049 (2019)}.

\bibitem{Medenjak2020} M. Medenjak, J. De Nardis and T. Yoshimura,
{Diffusion from convection}
\href{https://doi.org/10.21468/SciPostPhys.9.5.075}{SciPost Phys. {\bf9}, 075 (2020)}.

\bibitem{Durnin2021} J. Durnin, A. De Luca, J. De Nardis and B. Doyon,
{ Diffusive hydrodynamics of inhomogenous Hamiltonians},
\href{https://iopscience.iop.org/article/10.1088/1751-8121/ac2c57/meta}{J. Phys. A: Math. Theor. {\bf54}, 494001 (2021)}.

\bibitem{Bouchoule2020} I. Bouchoule, B. Doyon and J. Dubail,
{The effect of atom losses on the distribution of rapidities in the one-dimensional Bose gas },
\href{https://doi.org/10.21468/SciPostPhys.9.4.044}{SciPost Phys. {\bf9}, 044 (2020)}.

\bibitem{Durnin2020} J. Durnin, M.J. Bhaseen and B. Doyon,
Nonequilibrium Dynamics and Weakly Broken Integrability,
\href{https://journals.aps.org/prl/abstract/10.1103/PhysRevLett.127.130601}{Phys. Rev. Lett. {\bf127}, 130601 (2020)}.

\bibitem{Bastianello2020b} A. Bastianello, J. De Nardis and A. De Luca,
{ Generalized hydrodynamics with dephasing noise},
\href{https://journals.aps.org/prb/abstract/10.1103/PhysRevB.102.161110}{Phys. Rev. B {\bf 102},161110 (R) (2020)}.

\bibitem{Bastianello2021} A. Bastianello, A. De Luca and R. Vasseur,
{ Hydrodynamics of weak integrability breaking},
\href{https://iopscience.iop.org/article/10.1088/1742-5468/ac26b2/meta}{J Stat. Mech. (2021) 114003}.

\bibitem{Ruggiero2019} P. Ruggiero, Y. Brun, and J. Dubail,
 Conformal field theory on top of a breathing one-dimensional gas of hard core bosons,
\href{https://www.scipost.org/SciPostPhys.6.4.051/pdf}{SciPost Phys. {\bf 6}, 051 (2019)}.

\bibitem{Ruggiero2020} P. Ruggiero, P. Calabrese, B. Doyon and J. Dubail,
 Quantum Generalized Hydrodynamics,
\href{https://link.aps.org/doi/10.1103/PhysRevLett.124.140603}{Phys. Rev. Lett. {\bf124}, 140603 (2020)}.

\bibitem{Collura2020} M. Collura, A. De Luca, P. Calabrese and J. Dubail,
{Domain-wall melting in the spin-1/2 XXZ spin chain: emergent Luttinger liquid with fractal quasi-particle charge},
\href{https://doi.org/10.1103/PhysRevB.102.180409}{Phys. Rev. B 102, 180409(R) (2020)}.

\bibitem{Scopa2021a} S. Scopa, A. Krajenbrink, P. Calabrese and J. Dubail,
Exact entanglement growth of a one-dimensional hard-core quantum gas during a free expansion,
\href{https://iopscience.iop.org/article/10.1088/1751-8121/ac20ee/meta}{J. Phys. A: Math. Theor. {\bf54}, 404002 (2021)}.

\bibitem{Scopa2022b} F. Ares, S. Scopa and S. Wald,
{Entanglement dynamics of a hard-core quantum gas during a Joule expansion},
\href{https://doi.org/10.1088/1751-8121/ac8209}{J. Phys. A: Math. Theor. {\bf 55}, 375301 (2022)} .

\bibitem{Scopa2022c} S. Scopa and D.~X. Horvath,
Exact hydrodynamic description of symmetry-resolved R\'enyi entropies after a quantum quench,
\href{https://iopscience.iop.org/article/10.1088/1742-5468/ac85eb}{J. Stat. Mech. (2022) 083104}.

\bibitem{Rottoli2022} F. Rottoli, S. Scopa and P. Calabrese,
Entanglement Hamiltonian during a domain wall melting in the free Fermi chain,
\href{ https://doi.org/10.1088/1742-5468/ac72a1}{J. Stat. Mech. (2022) 063103}.

\bibitem{Ruggiero2022} P. Ruggiero, P. Calabrese, B. Doyon and J. Dubail,
Quantum generalized hydrodynamics of the Tonks–Girardeau gas: density fluctuations and entanglement entropy,
\href{https://iopscience.iop.org/article/10.1088/1751-8121/ac3d68/meta}{J. Phys. A: Math. Theor. {\bf55}, 024003 (2022)}.

\bibitem{Gochev1977} I. G. Gochev, Spin complexes in a bounded chain, \href{http://jetpletters.ru/ps/1375/article_20820.pdf}{JETP {\bf 26}, 3 (1977)}.

\bibitem{Gochev1983} I. G. Gochev, Contribution to the theory of plane domain walls in a ferromagnet, \href{http://www.jetp.ac.ru/cgi-bin/dn/e_058_01_0115.pdf}{JETP {\bf58}, 115 (1983)}.

\bibitem{Yuan2007} S.  Yuan,  H.  De  Raedt,  and  S.  Miyashita, Domain-wall dynamics near a quantum critical point,  \href{https://journals.aps.org/prb/abstract/10.1103/PhysRevB.75.184305}{Phys. Rev. B {\bf 75}, 184305 (2007)}.

\bibitem{Antal1999} T. Antal, Z. R\'acz, A. R\'akos and G. M. Sch\"utz,
 Transport in the XX chain at zero temperature: Emergence of flat magnetization profiles,
\href{https://journals.aps.org/pre/abstract/10.1103/PhysRevE.59.4912}{Phys. Rev. E {\bf59}, 4912 (1999)}.

\bibitem{Tasaki1} S. Tasaki,
Non-equilibrium stationary states of non-interacting electrons in a one-dimensional lattice,
\href{https://doi.org/10.1016/S0960-0779(01)00080-7}{Chaos, Solitons and Fractals {\bf12}, 2657 (2001)}.

\bibitem{Tasaki2} S. Tasaki,
Nonequilibrium stationary states for a quantum 1-d conductor,
\href{https://doi.org/10.1063/1.1291584}{AIP Conf. Proc. {\bf519}, 356 (2000)}.

\bibitem{Araki2000} H. Araki, T. G. Ho,
Asymptotic time evolution of a partitioned infinite two-sided isotropic XY- chain,
\href{https://www.mathnet.ru/links/3a28884c91c0b82f41b0b6d84a79ef2d/tm501.pdf}{Proc. Steklov Inst. Math. {\bf228}, 203 (2000)}.

\bibitem{Ogata2002} Y. Ogata,
Non-equilibrium properties in the transverse XX chain,
\href{https://journals.aps.org/pre/abstract/10.1103/PhysRevE.66.016135}{Phys. Rev. E {\bf66}, 016135 (2002)}.

\bibitem{Aschbacher2003} W. H. Aschbacher and C.~-A. Pillet,
Non-equilibrium steady states of the XY chain,
\href{https://doi.org/10.1023/A:1024619726273}{J. Stat. Phys. {\bf112}, 1153 (2003)}.

\bibitem{Antal2008} T. Antal, P. L. Krapivsky, and A. R\'akos,
 Logarithmic current fluctuations in nonequilibrium quantum spin chains,
\href{https://journals.aps.org/pre/abstract/10.1103/PhysRevE.78.061115}{Phys. Rev. E {\bf78}, 061115 (2008)}.

\bibitem{Doyon2015} B. Doyon, A. Lucas, K. Schalm, M. J. Bhaseen,
Non-equilibrium steady states in the Klein-Gordon theory,
\href{https://doi.org/10.1088/1751-8113/48/9/095002}{J. Phys. A {\bf48}, 095002 (2015)}.

\bibitem{Dubail2017} J. Dubail, J.-M. St\'ephan, J. Viti and P. Calabrese,
 Conformal field theory for inhomogeneous one-dimensional quantum systems: the example of non-interacting Fermi gases,
\href{https://scipost.org/SciPostPhys.2.1.002}{SciPost Phys. {\bf2}, 2 (2017)}.

\bibitem{Allegra2016} N. Allegra, J. Dubail, J.-M. St\'ephan and J. Viti,
 Inhomogeneous field theory inside the arctic circle,
\href{https://iopscience.iop.org/article/10.1088/1742-5468/2016/05/053108}{J. Stat. Mech. (2016) 053108}.

\bibitem{Karevski2002} D. Karevski, Scaling behaviour of the relaxation in quantum chains,
\href{https://link.springer.com/article/10.1140/epjb/e20020139}{Eur. Phys. J. B {\bf27}, 147 (2002)}.

\bibitem{Hunyadi2004} V.  Hunyadi,  Z.  R\`acz,  and  L.  Sasv\`ari, Dynamic scaling of fronts in the quantum XX chain,  
\href{https://journals.aps.org/pre/abstract/10.1103/PhysRevE.69.066103}{Phys. Rev. E {\bf 69}, 066103 (2004)}.

\bibitem{Rigol2004} M. Rigol and A. Muramatsu,
Emergence of Quasicondensates of Hard-Core Bosons at Finite Momentum,
\href{https://journals.aps.org/prl/abstract/10.1103/PhysRevLett.93.230404}{Phys. Rev. Lett. {\bf93}, 230404 (2004)}.

\bibitem{Rigol2015} L. Vidmar, J.~P. Ronzheimer, M. Schreiber, S. Braun, S.~S. Hodgman, S. Langer, F. Heidrich-Meisner, I. Bloch and U. Schneider,
Dynamical Quasicondensation of Hard-Core Bosons at Finite Momenta,
\href{https://journals.aps.org/prl/abstract/10.1103/PhysRevLett.115.175301}{Phys. Rev. Lett. {\bf115}, 175301 (2015)}.

\bibitem{Platini2005} T. Platini and D. Karevski, Scaling and front dynamics in Ising quantum chains,
\href{https://link.springer.com/article/10.11402Fepjb2Fe2005-00402-2}{Eur. Phys. J. B {\bf 48}, 225 (2005)}.

\bibitem{Platini2007} T. Platini and D. Karevski, Relaxation in the XX quantum chain,
\href{https://iopscience.iop.org/article/10.1088/1751-8113/40/8/002}{J. Phys. A {\bf40}, 1711 (2007)}.

\bibitem{DeLuca2013} A. De Luca, J. Viti, D. Bernard and B. Doyon, Nonequilibrium thermal transport in the quantum  Ising chain,
\href{https://journals.aps.org/prb/abstract/10.1103/PhysRevB.88.134301}{Phys. Rev. B {\bf88},1342301 (2013)}.

\bibitem{DeLuca2014} A. De Luca, G. Martelloni and J. Viti, Stationary states in a free fermionic chain from the quench action method,
\href{https://journals.aps.org/pra/abstract/10.1103/PhysRevA.91.021603}{Phys. Rev. A {\bf91}, 021603 (2014)}.

\bibitem{Viti2016} J.  Viti,  J.-M.  St\'ephan,  J.  Dubail  and  M.  Haque, Inhomogeneous quenches in a free fermionic  chain:  Exact results,
\href{https://iopscience.iop.org/article/10.1209/0295-5075/115/40011}{Europhysics Lett. {\bf115}, 40011 (2016)}.

\bibitem{Eisler2018} V. Eisler, F. Maislinger,
Hydrodynamical phase transition for domain-wall melting in the XY chain,
\href{https://doi.org/10.1103/PhysRevB.98.161117}{Phys. Rev. B {\bf98}, 161117(R) (2018)}.

\bibitem{Gobert2005} D.  Gobert,  C. Kollath,  U.  Schollw\"ock  and  G.  M.  Sch\"utz, 
Real-time dynamics in spin-$\frac{1}{2}$ chains with adaptive time-dependent density matrix renormalization group,
\href{https://journals.aps.org/pre/abstract/10.1103/PhysRevE.71.036102}{Phys. Rev. E {\bf71}, 036102 (2005)}.

\bibitem{Calabrese2008} P. Calabrese, C. Hagendorf and P. Le Doussal, Time evolution of 1D gapless models from a domain-wall initial state: stochastic Loewner evolution continued?,
\href{https://iopscience.iop.org/article/10.1088/1742-5468/2008/07/P07013}{J. Stat. Mech. (2008) P07013}.

\bibitem{Zauner2012} V. Zauner, M. Ganahl, H. Evertz, and T. Nishino, Time evolution within a comoving window: scaling of signal fronts and magnetization plateaus after a local quench in quantum spin chains, \href{https://doi.org/10.1088/0953-8984/27/42/425602}{J. Phys.: Cond. Matt. {\bf27}, 425602 (2012)}.

\bibitem{Halimeh2014} J. Halimeh, A. W\"ollert, I. Mc Culloch, U. Schollw\"ock and T.Barthel, Domain-wall melting in ultracold-boson systems with hole and spin-flip defects,
\href{https://journals.aps.org/pra/abstract/10.1103/PhysRevA.89.063603}{Phys. Rev. A {\bf89}, 063603 (2014)}.

\bibitem{Alba2014} V. Alba and F. Heidrich-Meisner,
 Entanglement spreading after a geometric quench in quantum spin chains,
\href{https://journals.aps.org/prb/pdf/10.1103/PhysRevB.90.075144}{Phys. Rev. B {\bf90}, 075144 (2014)}.

\bibitem{Vicari2012} E. Vicari,
 Quantum dynamics and entanglement in one-dimensional Fermi gases released from a trap,
\href{https://journals.aps.org/pra/abstract/10.1103/PhysRevA.85.062324}{Phys. Rev. A {\bf85}, 062324 (2012)}.

\bibitem{Sabetta2013} T.  Sabetta  and  G.  Misguich, Nonequilibrium steady states in the quantum XXZ spin chain, 
\href{https://journals.aps.org/prb/abstract/10.1103/PhysRevB.88.245114}{Phys. Rev. B {\bf88}, 245114 (2013)}.

\bibitem{Biella2016} A. Biella, A. De Luca, J. Viti, D. Rossini, L. Mazza and R. Fazio, Energy transport between  two integrable spin chains, 
\href{https://journals.aps.org/prb/abstract/10.1103/PhysRevB.93.205121}{Phys. Rev. B {\bf93}, 205121 (2016)}.

\bibitem{Bernard2016} D. Bernard and B. Doyon, Conformal field theory out of equilibrium: a review, 
\href{https://iopscience.iop.org/article/10.1088/1742-5468/2016/06/064005}{J. Stat. Mech. (2016) 064005}.

\bibitem{Vidmar2017} L. Vidmar, D. Iyer, and M. Rigol, Emergent Eigenstate Solution to Quantum Dynamics Far from Equilibrium, 
\href{https://journals.aps.org/prx/abstract/10.1103/PhysRevX.7.021012}{Phys. Rev. X {\bf 7}, 021012 (2017)}.

\bibitem{Langmann2017} E. Langmann, J. L. Lebowitz, V. Mastropietro and P. Moosavi,
Steady States and Universal Conductance in a Quenched Luttinger Model,
\href{https://link.springer.com/content/pdf/10.1007/s00220-016-2631-x.pdf}{Comm. Math. Phys. {\bf349}, 551 (2017)}.

\bibitem{abf-19} V. Alba, B. Bertini, and M. Fagotti, Entanglement evolution and generalised hydrodynamics: interacting integrable systems,
\href{https://doi.org/10.21468/SciPostPhys.7.1.005}{SciPost Phys. {\bf7}, 005 (2019)}.

\bibitem{bfpc-18} B. Bertini, M. Fagotti, L. Piroli, and P. Calabrese, {Entanglement evolution and generalised hydrodynamics: noninteracting systems},
\href{https://doi.org/10.1088/1751-8121/aad82e}{J. Phys. A  {\bf 51}, 39LT01 (2018)}.

\bibitem{mbpc-18} M. Mestyan, B. Bertini, L. Piroli, and P. Calabrese, {Spin-charge separation effects in the low-temperature transport of 1D Fermi gases},
\href{http://dx.doi.org/10.1103/PhysRevB.99.014305}{Phys. Rev. B {\bf 99}, 014305 (2019)}.

\bibitem{Jesenko2011} S. Jesenko  and  M.  Znidaric, Finite-temperature magnetization transport of the one-dimensional anisotropic Heisenberg model,
\href{https://journals.aps.org/prb/abstract/10.1103/PhysRevB.84.174438}{Phys. Rev. B{\bf84}, 174438 (2011)}.

\bibitem{Hauschild2016} J. Hauschild, F. Heidrich-Meisner and F. Pollmann, Domain-wall melting as a probe of many-body localization,
\href{https://journals.aps.org/prb/abstract/10.1103/PhysRevB.94.161109}{Phys. Rev.B {\bf94}, 161109(R) (2016)}.

\bibitem{Karrasch2013} C. Karrasch, R. Ilan, and J. E. Moore, Nonequilibrium thermal transport and its relation to linear response, 
\href{https://journals.aps.org/prb/abstract/10.1103/PhysRevB.88.195129}{Phys. Rev. B {\bf88},195129 (2013)}.

\bibitem{Rakovszky2019} T. Rakovszky, C. von Keyserlingk and F. Pollmann, Entanglement growth after inhomogenous quenches, \href{https://journals.aps.org/prb/abstract/10.1103/PhysRevB.100.125139}{Phys. Rev. B {\bf100},125139 (2019)}.

\bibitem{Lerose2020} A. Lerose, F. M. Surace, P. P. Mazza, G. Perfetto, M. Collura, and A. Gambassi,
Quasilocalized dynamics from confinement of quantum excitations,
\href{https://doi.org/10.1103/PhysRevB.102.041118}{Phys. Rev. B {\bf 102}, 041118(R) (2020)}.

\bibitem{Coppola2022} M. Coppola, G.~T. Landi, and D. Karevski,
Wigner dynamics for quantum gases under inhomogeneous gain and loss processes with dephasing,
\href{https://arxiv.org/abs/2212.11029}{preprint - arXiv: 2212.11029 (2022)}.

\bibitem{Misguich2017} G. Misguich, K. Mallick and P. L. Krapivsky, Dynamics of the spin-1/2 Heisenberg chain initialized in a domain-wall state, \href{https://journals.aps.org/prb/abstract/10.1103/PhysRevB.96.195151}{Phys. Rev. B {\bf 96}, 195151 (2017)}.

\bibitem{Misguich2019} G. Misguich, N. Pavloff and V. Pasquier, Domain  wall problem in the quantum XXZ chain and semiclassical behavior close to the isotropic point, \href{dx.doi.org/10.21468/SciPostPhys.7.2.025}{SciPost Phys. {\bf 7}, 025(2019)}.

\bibitem{Ljubotina2017} M. Ljubotina, M. Znidaric and T. Prosen, Spin diffusion from an inhomogeneous quench in an integrable system, \href{https://www.nature.com/articles/ncomms16117}{Nature Comm. {\bf 8}, 16117 (2017)}.

\bibitem{Ilievski2018} E. Ilievski, J. De Nardis, M. Medenjak and T. Prosen, Superdiffusion in one-dimensional quantum  lattice models, \href{https://journals.aps.org/prl/abstract/10.1103/PhysRevLett.121.230602}{Phys. Rev. Lett.{\bf121}, 230602 (2018)}.

\bibitem{Jordan1928} P. Jordan and E. Wigner, \"Uber das Paulische \"Aquivalenzverbot. 
\href{https://link.springer.com/article/10.1007/BF01331938}{Z. Phys. {\bf 47}, 631 (1928)}.

\bibitem{Wigner1997} E. P. Wigner,
{\it On the quantum correction for thermodynamic equilibrium},
\href{https://link.springer.com/chapter/10.1007/978-3-642-59033-7_9}{Springer-Verlag Berlin Heidelberg (1997)}.

\bibitem{Schutz1999} G. M. Sch\"utz and S. Trimper, 
Relaxation and aging in quantum spin systems,
\href{https://iopscience.iop.org/article/10.1209/epl/i1999-00367-8}{EPL {\bf 47}, 164 (1999)}.

\bibitem{Gradshteynp989} See e.g.~I. S. Gradshteyn and I. M. Ryzhik, {\it Table of Integrals, Series, and Products},  p.~989, Academic Press (1943).

\bibitem{Wendenbaum2013} P. Wendenbaum, M. Collura, and D. Karevski,
 Hydrodynamic description of hard-core bosons on a Galileo ramp,
\href{https://journals.aps.org/pra/abstract/10.1103/PhysRevA.87.023624}{Phys. Rev. A {\bf87}, 023624 (2013)}.

\bibitem{Fagotti2017} M. Fagotti,
Higher-order generalized hydrodynamics in one dimension: The noninteracting test,
\href{https://journals.aps.org/prb/abstract/10.1103/PhysRevB.96.220302}{Phys. Rev. B {\bf96}, 220302 (2017)}.

\bibitem{Fagotti2020} M. Fagotti,
Locally quasi-stationary states in noninteracting spin chains,
\href{https://scipost.org/10.21468/SciPostPhys.8.3.048}{SciPost Phys. {\bf8}, 048 (2020)}.

\bibitem{LR-bound} E. H. Lieb and D. W. Robinson, The Finite Group Velocity
of Quantum Spin Systems,
\href{https://doi.org/10.1007/BF01645779}{Commun. Math. Phys. {\bf28}, 251 (1972)}.

\bibitem{Takahashi-book} M. Takahashi, {\it Thermodynamics of one-dimensional solvable models} (Cambridge university press, 2005).

\bibitem{Korepin2010} V.~E. Korepin, N.~M. Bogoliubov, and A.~G. Izergin,
{\it  Quantum Inverse Scattering Method and Correlation Functions}, Cambridge Univ. Press (1993).

\bibitem{Doyon-geom} B. Doyon, H. Spohn and T. Yoshimura, 
{A geometric viewpoint on generalized hydrodynamics},
\href{https://doi.org/10.1016/j.nuclphysb.2017.12.002}{Nucl. Phys. B {\bf926}, 570 (2017)}.

\bibitem{Pozgay2020} B. Pozsgay,
Algebraic Construction of Current Operators in Integrable Spin Chains,
\href{https://doi.org/10.1103/PhysRevLett.125.070602}{Phys. Rev. Lett. {\bf125}, 070602 (2020)}.

\bibitem{Pozgay2020b} M. Borsi, B. Pozsgay and L. Pristy\`ak,
Current Operators in Bethe Ansatz and Generalized Hydrodynamics: An Exact Quantum-Classical Correspondence,
\href{https://doi.org/10.1103/PhysRevX.10.011054}{Phys. Rev. X {\bf10}, 011054 (2020)}.

\bibitem{Borsl2021} M. Borsi, B. Pozsgay and L. Pristy\`ak,
Current operators in integrable models: a review,
\href{https://iopscience.iop.org/article/10.1088/1742-5468/ac0f6b/meta}{J. Stat. Mech. (2021) 094001}.

\bibitem{iTensor} M. Fishman, S. R. White, E. M. Stoudenmire, The iTensor Software Library for Tensor Network Calculations, \href{https://arxiv.org/pdf/2007.14822.pdf}{preprint - arXiv: 2007.14822 (2020)}.

\bibitem{p-12} I. Peschel,  Entanglement in solvable many-particle models,
 \href{http://dx.doi.org/10.1007/s13538-012-0074-1}{Braz. J. Phys. {\bf 42}, 267 (2012)}.

\bibitem{Peschel1999a} I. Peschel, M. Kaulke and \"O. Legeza,
 Density-matrix spectra for integrable models,
\href{https://doi.org/10.1002/andp.19995110203}{Ann. Phys. {\bf8}, 153 (1999)}.

\bibitem{Chung2001} M. C. Chung and I. Peschel,
 Density-matrix spectra of solvable fermionic systems,
\href{https://journals.aps.org/prb/abstract/10.1103/PhysRevB.64.064412}{Phys. Rev. B {\bf64}, 064412 (2001)}.

\bibitem{Peschel2003} I. Peschel,
 Calculation of reduced density matrices from correlation functions,
\href{https://iopscience.iop.org/article/10.1088/0305-4470/36/14/101}{J. Phys. A {\bf36}, L205 (2003)}.

\bibitem{Peschel2004} I. Peschel,
 On the reduced density matrix for a chain of free electrons,
\href{https://iopscience.iop.org/article/10.1088/1742-5468/2004/06/P06004}{J. Stat. Mech. (2004) P06004}.

\bibitem{Peschel2009} I. Peschel and V. Eisler,
 Reduced density matrices and entanglement entropy in free lattice models,
\href{https://iopscience.iop.org/article/10.1088/1751-8113/42/50/504003}{J. Phys. A {\bf42}, 504003 (2009)}.

\bibitem{Eisler2008} V. Eisler, D. Karevski, T. Platini and I. Peschel,
Entanglement evolution after connecting finite to infinite quantum chains,
\href{https://iopscience.iop.org/article/10.1088/1742-5468/2008/01/P01023}{J. Stat. Mech. (2008) P0102}.

\bibitem{Eisler2014} V. Eisler and I. Peschel,
Surface and bulk entanglement in free-fermion chains,
\href{https://iopscience.iop.org/article/10.1088/1742-5468/2014/04/P04005/meta}{J. Stat. Mech. (2014) P04005}.

\bibitem{Eisler2016} V. Eisler, F. Maislinger and H. G. Evertz,
Universal front propagation in the quantum Ising chain with domain-wall initial states,
\href{https://scipost.org/SciPostPhys.1.2.014/pdf}{SciPost Phys. {\bf1}, 014 (2016)}.

\bibitem{Eisler2017} V. Eisler and I. Peschel,
Analytical results for the entanglement hamiltonian of a free-fermion chain,
\href{https://iopscience.iop.org/article/10.1088/1751-8121/aa76b5}{J. Phys. A {\bf50}, 284003 (2017)}.

\bibitem{Eisler2021} V. Eisler, 
Entanglement spreading after local and extended excitations in a free-fermion chain,
To be published in \href{https://doi.org/10.1088/1751-8121/ac21e4}{J. Phys. A: Math. Theor. (2021)}.

\bibitem{Giamarchi2007} T. Giamarchi,
%\href{https://oxford.universitypressscholarship.com/view/10.1093/acprof:oso/9780198525004.001.0001/acprof-9780198525004}{
{{\it Quantum Physics in One Dimension}, Oxford Univ. Press (2007)}.

\bibitem{Cazalilla2004} M. A. Cazalilla,
 Bosonizing one-dimensional cold atomic gases,
\href{https://iopscience.iop.org/article/10.1088/0953-4075/37/7/051}{J. Phys. B {\bf 37}, 1 (2004)}.

\bibitem{Cazalilla2011} M. A. Cazalilla, R. Citro, T. Giamarchi, E. Orignac, and M. Rigol,
One dimensional bosons: From condensed matter systems to ultracold gases,
\href{https://journals.aps.org/rmp/abstract/10.1103/RevModPhys.83.1405}{Rev. Mod. Phys. {\bf 83}, 1405 (2011)}.

\bibitem{Brun2017} Y. Brun and J. Dubail,
 One-particle density matrix of trapped one-dimensional impenetrable bosons from conformal invariance,
\href{https://scipost.org/SciPostPhys.2.2.012/pdf}{SciPost Phys. {\bf2}, 012 (2017)}.

\bibitem{Brun2018} Y. Brun and J. Dubail,
 The Inhomogeneous Gaussian Free Field, with application to ground state correlations of trapped 1d Bose gases, 
\href{https://www.scipost.org/SciPostPhys.4.6.037/pdf}{SciPost Phys. {\bf4}, 037 (2018)}.

\bibitem{Bastianello2020} A. Bastianello, J. Dubail, and J.-M. St\'ephan,
 Entanglement entropies of inhomogeneous Luttinger liquids,
\href{https://doi.org/10.1088/1751-8121/ab7580}{J. Phys. A {\bf 53}, 23 (2020)}.

\bibitem{Scopa2020} S. Scopa, L. Piroli and P. Calabrese,
 One-particle density matrix of a trapped Lieb Liniger anyonic gas, 
\href{https://iopscience.iop.org/article/10.1088/1742-5468/abaed1}{J. Stat. Mech. (2020) 093103}.

\bibitem{Calabrese2004} P. Calabrese and J. Cardy
Entanglement entropy and quantum field theory,
\href{https://iopscience.iop.org/article/10.1088/1742-5468/2004/06/P06002}{J. Stat. Mech. (2004) P06002}.

\bibitem{Cardy2008} J. L. Cardy, O. A. Castro-Alvaredo and B. Doyon,
 Form factors of branch-point twist fields in quantum integrable models and entanglement entropy,
\href{https://link.springer.com/article/10.1007/s10955-007-9422-x}{J. Stat. Phys. {\bf130}, 129 (2008)}.

\bibitem{Calabrese2009} P. Calabrese and J. Cardy,
 Entanglement entropy and conformal field theory,
\href{https://iopscience.iop.org/article/10.1088/1751-8113/42/50/504005}{J. Phys. A {\bf 42}, 504005 (2009)}.

\bibitem{Calabrese2010} P. Calabrese and F. H.L. Essler,
 Universal corrections to scaling for block entanglement in spin-1/2 XX chains,
\href{https://iopscience.iop.org/article/10.1088/1742-5468/2010/08/P08029}{J. Stat. Mech. (2010) P08029}.

\bibitem{Jin2004} B. Q. Jin and V. E. Korepin,
 Quantum spin chain, Toeplitz determinants and the Fisher-Hartwig conjecture,
\href{https://link.springer.com/article/10.1023/B:JOSS.0000037230.37166.42}{J. Stat. Phys. {\bf116}, 79 (2004)}.

\bibitem{Holzhey-Larsen-Wilczek} C. Holzhey, F. Larsen and F. Wilczek,
{Geometric and renormalized entropy in conformal field theory},
\href{https://doi.org/10.1016/0550-3213(94)90402-2}{Nucl. Phys. {\bf B424}, 443--467 (1994)}.


\end{thebibliography}
